\documentclass[prd,aps,twocolumn,nofootinbib,preprintnumbers,superscriptaddress,preprintnumbers,balancelastpage,longbibliography]{revtex4-2}
\usepackage[english]{babel}

\usepackage{amsmath,mathtools,physics,xfrac}
\usepackage{graphicx}
\usepackage{afterpage}
\usepackage{float}
\usepackage{subfigure}
\usepackage{rotating}
\usepackage{multirow}
\usepackage{tabularx}
\usepackage{booktabs}
\usepackage{fancyhdr}
\usepackage[utf8]{inputenc}
\usepackage{theorem}
\usepackage{moreverb}
\usepackage{euscript}
\usepackage{psfrag}
\usepackage{slashed}
\usepackage{mathtools}
\usepackage{makecell}
\usepackage{adjustbox}
\usepackage{dcolumn}
\usepackage{bm}
\usepackage[dvipsnames]{xcolor}
\usepackage{graphics}
\usepackage{hyperref}
\usepackage{chngcntr}
\usepackage{aas_macros}
\usepackage{tabularx}
\usepackage{color,xcolor}
\usepackage[normalem]{ulem}

\hypersetup{
     colorlinks   = true,
     citecolor    = Aquamarine,
     urlcolor     = Aquamarine,
     linkcolor    = Aquamarine
}

\definecolor{darkblue}{rgb}{0.0,0.0,0.75}
\definecolor{darkred}{rgb}{0.6,0.0,0}
\definecolor{darkgreen}{rgb}{0.0,0.6,0.}

\counterwithout{equation}{section}

\usepackage{tikz,xcolor,hyperref}

\definecolor{lime}{HTML}{A6CE39}
\DeclareRobustCommand{\orcidicon}{\hspace{-1mm}
	\begin{tikzpicture}
		\draw[lime, fill=lime] (0,0) 
		circle [radius=0.16] 
		node[white] {{\fontfamily{qag}\selectfont \tiny \,ID}};
		\draw[white, fill=white] (-0.0525,0.095) 
		circle [radius=0.007];
	\end{tikzpicture}
	\hspace{-3mm}
}

\foreach \x in {A, ..., Z}{\expandafter\xdef\csname orcid\x\endcsname{\noexpand\href{https://orcid.org/\csname orcidauthor\x\endcsname}
		{\noexpand\orcidicon}}
}


\begin{document}

\title{\boldmath Breaking Dark: Hunting Heavy Decaying Dark Matter with Tibet AS$_{\gamma}$ and LHAASO-KM2A}

\author{Abhishek Dubey\orcidA{}}
\email{abhishekd1@iisc.ac.in}
\affiliation{Centre for High Energy Physics, Indian Institute of Science, C.\,V.\,Raman Avenue, Bengaluru 560012, India}

\author{Akash Kumar Saha\orcidB{}}
\email{akashks@iisc.ac.in}

\affiliation{Centre for High Energy Physics, Indian Institute of Science, C.\,V.\,Raman Avenue, Bengaluru 560012, India}

	
	
	\begin{abstract}

Recent measurements of diffuse sub-PeV gamma rays by the Tibet AS$_\gamma$ and LHAASO collaborations have reshaped our understanding of the gamma-ray sky. Besides uncovering the nature of `PeVatrons', these measurements can also be used to probe the non-gravitational nature of dark matter. PeV-scale decaying dark matter can produce high-energy gamma rays in the final state and contribute to the measurements made by extensive air-shower detectors like Tibet AS$_\gamma$ and LHAASO. Using the latest Tibet AS$_\gamma$ upper limits on diffuse gamma rays away from the Galactic plane and the LHAASO-KM2A measurements of diffuse gamma rays from the Galactic plane, we put stringent constraints on lifetimes of decaying DM for masses $\sim 10^6 - 10^9$ GeV. Future observations of high-energy diffuse gamma-ray emission can thus provide stronger limits or potentially discover heavy decaying dark matter.

	\end{abstract}
	
	\maketitle

\section{\label{sec:level1}Introduction}

Dark matter (DM), constituting approximately 27\% of the Universe's energy content, remains one of modern physics' most profound mysteries\,\cite{Bertone:2016nfn}. While its gravitational effects are well-established through multiple astronomical and cosmological observations---including galactic rotation curves\,\cite{Rubin:1970zza}, gravitational lensing\,\cite{Clowe:2006eq}, and cosmic microwave background anisotropies\,\cite{Planck:2018vyg}---its particle nature and non-gravitational interactions continue to elude detection\,\cite{Cirelli:2024ssz, Strigari:2012acq, Lisanti:2016jxe, Slatyer:2017sev, Lin:2019uvt}. The mass range of potential DM candidates spans many orders of magnitude, with the PeV scale being particularly intriguing from various theoretical considerations\,\cite{Chung:1998zb, Chung:1998rq, Giudice:1999yt, Chung:1998ua, Steffen:2006hw,Harigaya:2013pla,Harigaya:2014waa,Garny:2015sjg,Dev:2016qbd,
Berlin:2016vnh,Harigaya:2016nlg, Babichev:2018mtd, Kim:2019udq,Dudas:2020sbq,Hambye:2020lvy,Mambrini:2022uol,Giudice:2024tcp,Carney:2022gse,Deligny:2024fyx}.

\begin{figure}[t]
\centering
\includegraphics[width=\columnwidth]{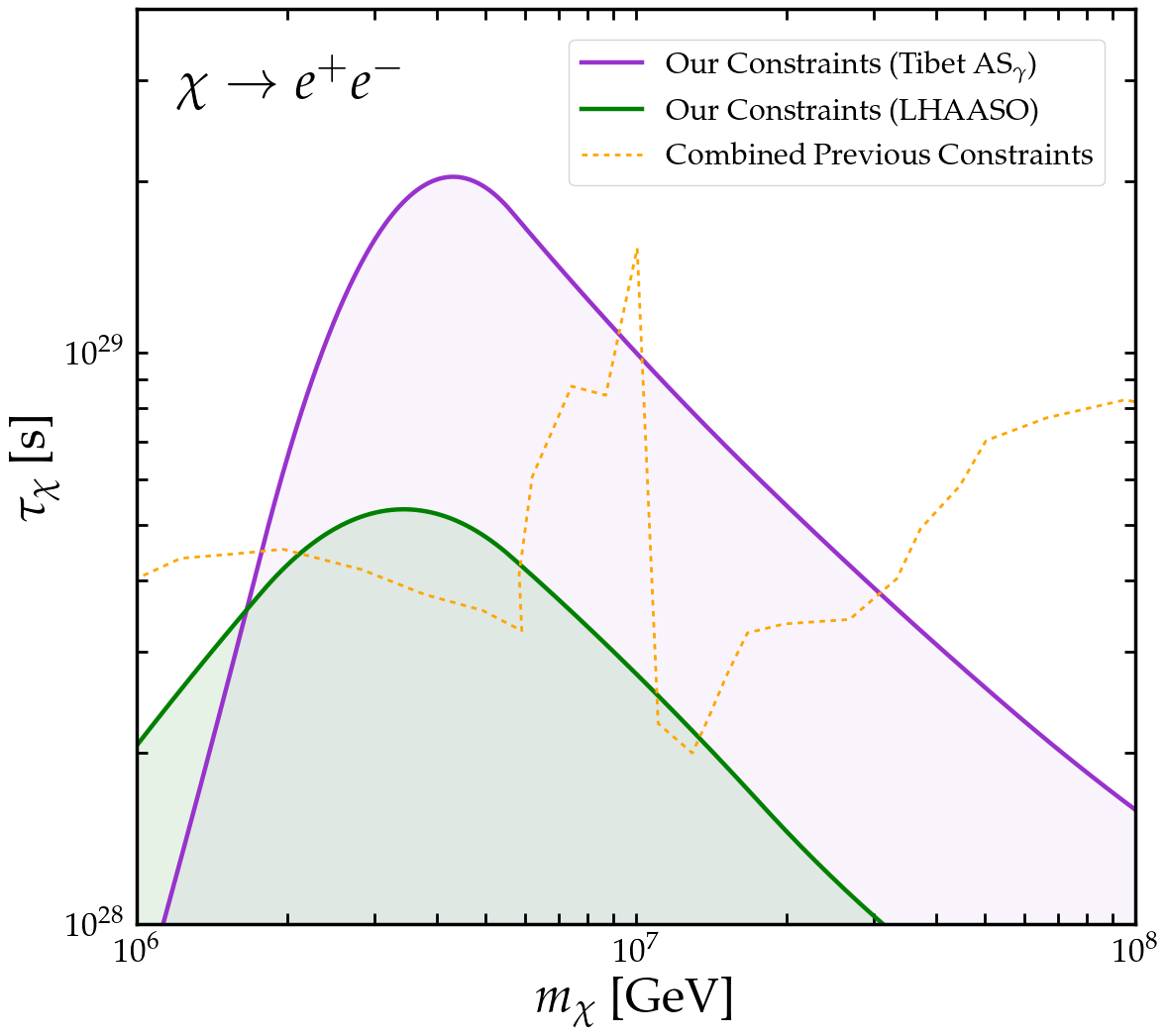}
\caption{Upper limits on DM lifetime, $\tau_\chi$, as a function of its mass $m_\chi$ for the decay channel $\chi \to e^{+} e^{-}$. Our limits from Tibet AS$_\gamma$ (Neronov \textit{et al.})\,\cite{Neronov:2021ezg} and LHAASO-KM2A\,\cite{LHAASO:2023gne} datasets are shown by the purple solid and green solid lines, respectively. The excluded regions lie below the
curves. Previous combined best bounds in the parameter space are taken from Refs.\,\cite{Esmaili:2014rma,Cohen:2016uyg,Blanco:2018esa,IceCube:2018tkk,Bhattacharya:2019ucd,Ishiwata:2019aet,Chianese:2019kyl,LHAASO:2022yxw,LHAASO:2024upb} (orange dashed line).  }
\label{fig:ee_constraints}
\end{figure}

High-energy gamma rays can be produced either via hadronic or leptonic processes due to cosmic rays interacting with the interstellar medium or near astrophysical sources\,\cite{Moskalenko:2004vh, Kelner:2006tc,Kappes:2006fg,Abdo:2007ad,2009herb.book.....D,2012APh....35..503G,Lipari:2018gzn,Moskalenko:2005ng,1989ApJ...342..379W, Capanema:2020oet, 2010ApJ...710.1530V,Vernetto:2016alq,DeAngelis:2013jna,Ruffini:2015oha,Sudoh:2022sdk,1971NASSP.249.....S, Bose:2022ghr, Cardillo:2023hbb,Lipari:2024pzo,Ehlert:2023btz}. Recent advancements in ground-based gamma-ray astronomy, particularly through the Tibet AS$_\gamma$ and Large High Altitude Air Shower Observatory (LHAASO), have enabled the detection of gamma rays in the TeV-PeV energy range. The Tibet AS$_\gamma$ collaboration has reported observations of diffuse gamma rays in the energy range of  100 TeV to 1 PeV, from two distinct Galactic plane regions: $25^\circ < l < 100^\circ$, $|b| < 5^\circ$ and $50^\circ < l < 200^\circ$, $|b| < 5^\circ$, where $l$ and $b$ denote Galactic longitude and latitude, respectively\,\cite{TibetASgamma:2021tpz}. More recently, LHAASO has reported detections of diffuse gamma-ray emission in the energy range of 10 TeV to 1 PeV using its Square Kilometer Array (KM2A). The measurements reveal signals from the Galactic plane in both the inner galaxy ($15^\circ < l < 125^\circ$, $|b| < 5^\circ$) and outer galaxy ($125^\circ < l < 235^\circ$, $|b| < 5^\circ$) regions, with significances of $29.1\sigma$ and $12.7\sigma$ respectively\,\cite{LHAASO:2023gne}. Besides direct observations of gamma rays from the Galactic plane, the all particle cosmic-ray detection by Tibet AS$_\gamma$ has been used to derive stringent upper limits on diffuse gamma-ray flux away from the Galactic plane\,\cite{Neronov:2021ezg}.  These detections are revolutionizing our understanding of the cosmic-ray flux and the location as well as the nature of the nearby cosmic-ray accelerators at the highest energies.

Indirect detection of DM, which searches for DM decay or annihilation to Standard Model (SM) particles, represents a promising avenue for identifying its fundamental nature. Among the various possible decay products, gamma rays can be produced by decay or hadronization of these SM particles and are particularly valuable probes due to their ability to travel astronomical distances without significant deflection, carrying pristine information about their origin. These gamma-rays can be produced through primary emission and secondary processes. Prompt gamma-rays emerge directly from the decay of DM particles into SM final states through hadronization, particle decay, and electroweak processes\,\cite{Cirelli:2010xx, Bauer:2020jay,Arina:2023eic}. Secondary emissions arise when high-energy electrons and positrons from DM decay undergo inverse Compton (IC) scattering with low-energy background photons, including the cosmic microwave background (CMB), infrared radiation (IR), starlight (SL), and extragalactic background light (EBL).

\begin{figure*}
	\begin{center}
		\includegraphics[width=\columnwidth]{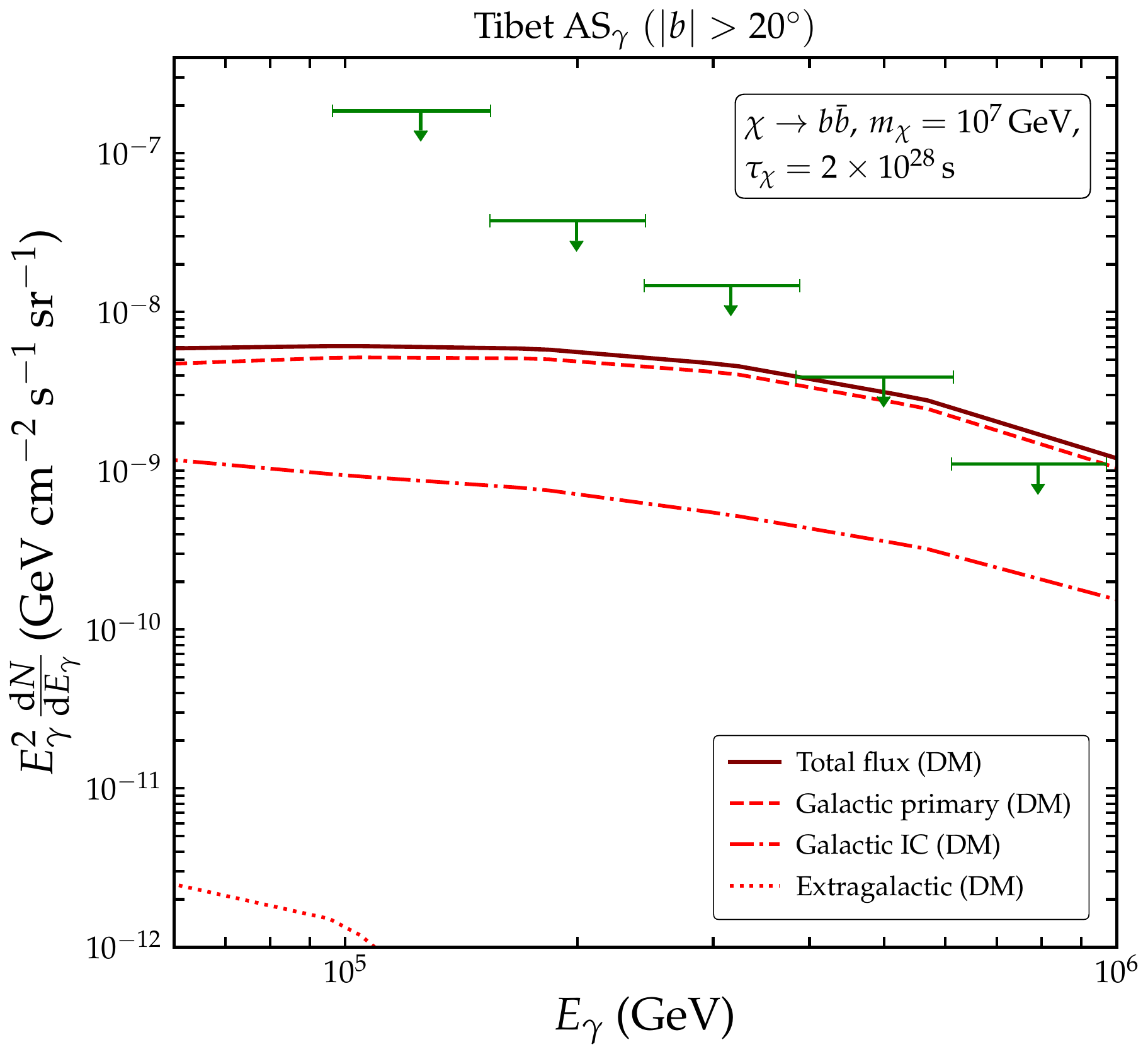}~~
		\includegraphics[width=\columnwidth]{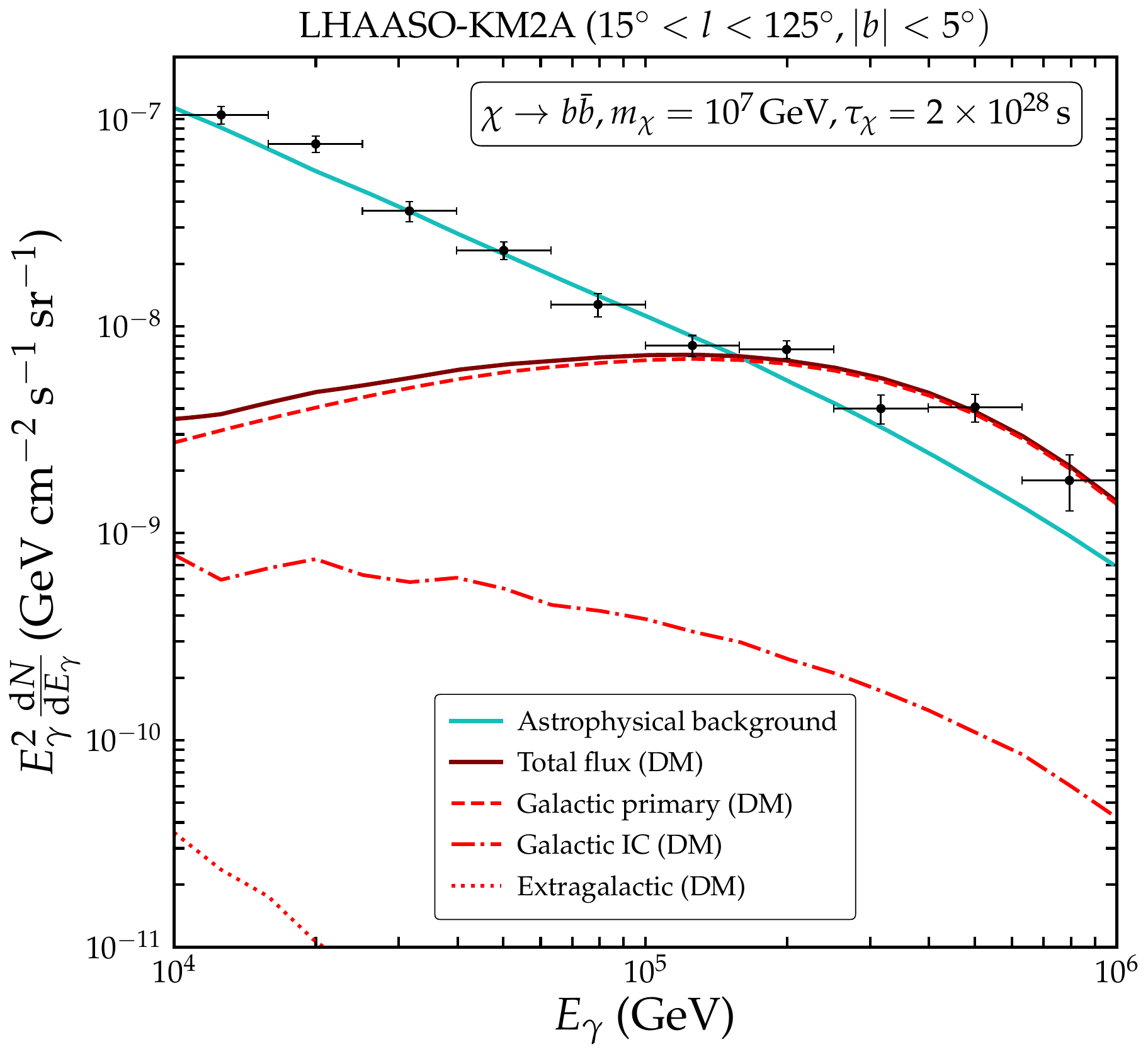}~~\\	
		\caption{ Differential gamma-ray flux as a function of photon energies for the DM decay channel $\chi \to b\bar{b}$ with $m_{\chi} = 10^{7}\,\mathrm{GeV}$ and $\tau_{\chi} = 2 \times 10^{28}\,\mathrm{s}$, along with the Tibet AS$_\gamma$ upper limits (green upper limits in the \textbf{left panel})\,\cite{Neronov:2021ezg} and LHAASO-KM2A inner galaxy datasets (black data points in the \textbf{right panel})\,\cite{LHAASO:2023gne}. In both the panels, the total flux from DM decay (maroon solid line) is shown along with contributions from Galactic primary emission (red dashed line), Galactic IC (red dot-dashed line), and extragalactic components (red dotted line). In the right panel, the LHAASO-KM2A measurements are overlaid together with the modeled astrophysical background from Chen \textit{et al.}\,\cite{Chen:2024yin}. The chosen mass and lifetime parameters are excluded by our analysis (see Fig.\,\ref{fig: bbbar_constraints}). Note that the $x$ and $y$ axis range in two panels are not the same.}
		\label{fig:flux_plots}
	\end{center}	
\end{figure*}

In this work, we analyze the upper limits from Tibet AS$_\gamma$ and flux measurements from LHAASO-KM2A to constrain PeV-scale decaying DM models. We consider a wide range of DM decay channels into SM particles, including quarks, leptons, and gauge bosons, accounting for both prompt and IC emission components. Previously, high energy cosmic-ray, gamma-ray, and neutrino observations have been used to put stringent constraints on heavy decaying DM\,\cite{Ishiwata:2008cu, Murase:2012xs, Murase:2015gea, Esmaili:2015xpa, Esmaili:2014rma,Murase:2015gea, Kalashev:2016cre,Kalashev:2017ijd, Kachelriess:2018rty, Blanco:2018esa, Ishiwata:2019aet, Alcantara:2019sco, Cohen:2016uyg, Bhattacharya:2019ucd,Kalashev:2019xkw,  Kuznetsov:2016fjt,Sui:2018bbh,Chianese:2019kyl,Maity:2021umk,Anchordoqui:2021crl, Esmaili:2021yaw,Chianese:2021jke,IceCube:2022clp,LHAASO:2022yxw,Arguelles:2022nbl,Aramaki:2022zpw,Skrzypek:2022hpy,Hambye:2021moy,Allahverdi:2023nov,IceCube:2023gku,PierreAuger:2023vql,IceCube:2023ies, PierreAuger:2023vql, LHAASO:2024upb,Sarmah:2024ffy,Das:2024bed,Dekker:2019gpe,Ng:2020ghe,Deliyergiyev:2022bvp,Leung:2023gwp,Das:2023wtk,Munbodh:2024ast,Song:2024vdc,Liu:2025vpz,Murase:2025uwv,Berghaus:2025jwb,LAT:2025wdk}. We note that the recent observation of the highest energy neutrino event at KM3NeT has sparked interest for both interpretation and bounds on heavy DM\,\cite{Borah:2025igh,Khan:2025gxs,Murase:2025uwv, Barman:2025hoz,Jho:2025gaf,Jiang:2025blz,Kohri:2025bsn,Aloisio:2025nts}. Future telescopes will offer improved sensitivities for heavy DM searches\,\cite{Ng:2020ghe, Rinchiuso:2020skh,Montanari:2022buj,Rodd:2024qsi,Chianese:2021htv,Zhu:2023bex,Baumgart:2025dov,Abe:2025qsi,Aghaie:2025iyn}. Our analysis demonstrates that the recent Tibet AS$_\gamma$ and LHAASO observations provide the most stringent constraints to date on DM lifetime for several decay channels in the PeV mass range, surpassing previous bounds from complementary observations, including IceCube, KM3NeT, LHAASO and Fermi-LAT\,\cite{Cohen:2016uyg, Aloisio:2025nts, IceCube:2018tkk,Ishiwata:2019aet, Chianese:2019kyl,Maity:2021umk, Esmaili:2021yaw, LHAASO:2022yxw,LHAASO:2024upb}.

In Fig.\,\ref{fig:ee_constraints} we show our limits for the DM decay channel, $\chi \to e^{+}e^{-}$ using the Tibet AS$_\gamma$ (purple solid line) and LHAASO (green solid line) measurements. The combined previous limits are shown by the orange dashed line. Our analysis thus provides the most stringent constraints in a significant part of the DM mass range.

\section{Gamma-Ray Flux from Dark Matter Decay}
\label{DM decay}

The gamma-ray flux from DM decay consists of four components: galactic prompt, IC emission, and their extragalactic counterparts. Each of these components shows distinct spectral behavior depending on the final states under consideration. We present a detailed discussion of each component below.

\subsection{Galactic Prompt Emission}
The prompt gamma-ray flux from DM decay is produced by the final-state SM particles. For example, in the $\chi \to b\bar{b}$ channel, gamma rays are produced through hadronization and subsequent particle decays. Even for leptonic channels like $\chi \to e^+e^-$, prompt gamma rays can be produced through electroweak corrections\,\cite{Bell:2008ey, Kachelriess:2009zy, Cirelli:2010xx, Ciafaloni:2010ti, Bauer:2020jay}.

The differential gamma-ray flux from prompt emission is given by
\begin{equation}
   \frac{d \phi^{G}_{\rm prompt}}{d E_\gamma d \Omega}=\frac{1}{4 \pi m_\chi \tau_\chi} \frac{d N_\gamma}{d E_\gamma} \int_0^{\infty} d s \,\rho(s, b, l) \,e^{-\tau_{\gamma \gamma}\left(E_\gamma, s, b, l\right)}\,\,,
\end{equation}
where $m_{\chi}$ is the DM mass, $\tau_\chi$ is the DM decay lifetime, $dN_\gamma/dE_\gamma$ is the gamma-ray spectrum per DM decay to various final states, $s$ is the line-of-sight distance, and $\tau_{\gamma \gamma}$ is the optical depth due to CMB, SL, and IR. We use the publicly available code \texttt{HDMSpectra}\,\cite{Bauer:2020jay} to calculate $dN_\gamma/dE_\gamma$ for DM decay to various SM final states.

For the DM density profile in the Milky Way, we assume the NFW profile\,\cite{Navarro:1996gj}
\begin{equation}
    \rho(r) = \frac{\rho_s}{(r/r_s)(1 + r/r_s)^2}\,\,,
\end{equation}
with $r_s = 20$ kpc and $\rho_s = 0.318$ GeV cm$^{-3}$. The distance $r$ from the Galactic centre is
\begin{equation}
    r(s,b,l) = \sqrt{R_{\odot}^2 - 2sR_{\odot}\cos(b) \cos(l) + s^2}\,\,,
\end{equation}
where $R_\odot = 8.3$ kpc is the Sun's distance from the Galactic centre\,\cite{2023MNRAS.519..948L}.

\subsection{Galactic Inverse Compton Emission}
Electrons and positrons produced from various SM final states in DM decay can upscatter background photons to gamma-ray energies through IC scattering. The background photon fields include CMB, SL, and IR.

The IC gamma-ray flux is given by\,\cite{ Cirelli:2009IC, Meade:2009iu, Cirelli:2010xx,Delahaye:2007fr, Esmaili:2015xpa} 
\begin{equation}
\begin{aligned}
\frac{d \phi^G_{\mathrm{IC} }}{d E_\gamma d \Omega} &= 
\frac{2}{E_\gamma} \cdot \frac{1}{4 \pi m_{\chi} \tau_{\chi}} \int_{m_e}^{m_{\chi}/2} d E_{\mathrm{s}} \, \frac{d N_{e}}{d E_e}\left(E_{\mathrm{s}}\right) \\
&\times \int_{\text{l.o.s.}} d s \, e^{-\tau_{\gamma \gamma}\left(E_\gamma, s, b, l\right)} \, \rho(s, b, l) \\ 
&\times \int_{m_e}^{E_{\mathrm{s}}} d E_e \, \frac{\sum_i \mathcal{P}_{\mathrm{IC}}^i\left( E_\gamma, E_e, s, b, l \right)}{b^G_{\rm tot}(E_e, s, b, l)}  I_{\text{diff}}\left(E_e, E_{\mathrm{s}}, s, b, l\right)\,\,,
\end{aligned}
\end{equation} 
where $E_e$ is the prompt electron-positron energy, $b^G_{\rm tot}$ is the energy loss parameter, $\sum_i \mathcal{P}_{\mathrm{IC}}^i$ represents the differential power emitted as photons through the ICS, and $I_{\mathrm{diff}}$ is the diffusion halo function. The latter summation runs over different photon bath components, including the CMB, SL, and IR. Since the electrons relevant here reach multi-TeV energies, we compute the IC term using the full Klein-Nishina kernel throughout. The Thomson limit forms  are adequate only for the lowest-energy CMB-scattering regime\,\cite{Cirelli:2009IC}.

For high-energy electrons and positrons ($e^\pm$) propagating through the Galaxy, two processes dominate their energy losses: synchrotron radiation from the Galactic magnetic field and inverse Compton (IC) scattering off the ambient photon backgrounds. Because these fields vary with position, the loss rate $b_{\rm tot}(E_e,\bar{x})$ is location dependent. Accordingly, the total energy loss function is
\begin{eqnarray}
    b^G_{\rm tot}(E_e,\bar{x}) \equiv -\frac{dE_e}{dt} = b^G_{\mathrm{IC}}(E_e,\bar{x}) + b^G_{\mathrm{syn}}(E_e,\bar{x})\,,
\end{eqnarray}
where the terms on the right hand side are energy loss rates due to IC scattering and synchrotron, respectively.

For the IC term, we employ the full Klein–Nishina expression obtained by integrating over the local radiation field\,\cite{Esmaili:2015xpa}

\begin{align}
b^G_{\mathrm{IC}}(E_e,\bar{x})
&= 3\,\sigma_T \int_0^\infty d\varepsilon\,\varepsilon
   \int_{1/(4\gamma^2)}^{1} dq\; n(\varepsilon,\bar{x}) \nonumber
   \frac{[(4\gamma^2-\Gamma_\varepsilon)q-1]}{(1+\Gamma_\varepsilon q)^3}\\
&\quad\times
   \left[
      2q\ln q + q + 1 - 2q^2
      + \frac{(\Gamma_\varepsilon q)^2(1-q)}{2(1+\Gamma_\varepsilon q)}
   \right]\,\,,
\end{align}
with $\gamma=E_e/m_e$ and $\Gamma_\varepsilon=4\varepsilon\,\gamma/m_e$. Here $\sigma_T$ is the Thomson cross section and $n(\varepsilon,\bar{x})$ is the differential number density of SL\,+\,IR\,+\,CMB photons with energy $\varepsilon$ at position $\bar{x}$\,\cite{Vernetto:2016alq}.

Synchrotron losses follow from the magnetic-energy density at $\bar{x}$:
$$
b^G_{\mathrm{syn}}(E_e,\bar{x})
= \frac{4\,\sigma_T\,E_e^2}{3\,m_e^2}\;\frac{B^2(\bar{x})}{2}\,.
$$
 The Galactic magnetic field consists of a  regular and a
halo/ turbulent component. For the regular magnetic field, we consider the spatial profile given in Ref.\,\cite{Strong:2000Diffuse}
\begin{equation}
B_{\mathrm{reg}}(\vec{x}) \equiv B_{\mathrm{reg}}(r, z) =
B_{0} \exp\!\left(-\frac{|r - R_{\odot}|}{r_{B}} - \frac{|z|}{z_{B}}\right),
\label{eq:Bfield_profile}
\end{equation}
where $R_{\odot} = 8.3~\mathrm{kpc}$, $r_{B} = 10~\mathrm{kpc}$, 
$z_{B} = 2~\mathrm{kpc}$, and $B_{0} = 4.78~\mu\mathrm{G}$. We assume a uniform constant strength magnetic field for the halo component which can be of the order of $\sim 1 \,\mu$G\,\cite{Esmaili:2015xpa}. 

The IC differential power emitted for $i^{\rm th}$ background photon component is given as\,\cite{Cirelli:2010xx}
\begin{equation}
\begin{aligned}
\mathcal{P}_{\mathrm{IC}}^i(E_\gamma, E_e, \vec{x}) 
&= \frac{3\sigma_T}{4\gamma^2} 
\int_{1/(4\gamma^2)}^1 \frac{dq}{q} \,
\Bigg[
   E_\gamma - \frac{E_\gamma}{4 q \gamma^2 (1 - \epsilon)}
\Bigg] \\
&\quad \times n_i(E_\gamma^0(q), \vec{x}) \\
&\quad \times \Bigg[
   2q \ln q + q + 1 - 2q^2
   + \tfrac{1}{2} \frac{\epsilon^2}{(1 - \epsilon)} (1 - q)
\Bigg],
\end{aligned}
\end{equation}

with,
\begin{equation}
\begin{aligned}
    q &= \frac{\epsilon}{\Gamma_E (1 - \epsilon)},\, 
    \Gamma_E = \frac{4 E_\gamma^0 E_e}{m_e^2},\, 
    \epsilon = \frac{E_\gamma}{E_e}, \\
    &\text{for} \quad \frac{1}{4\gamma^2} \leq q \leq 1.
\end{aligned}
\end{equation}
Here, $E_\gamma^0$ is the initial energy of the photon in the background bath. Accordingly, $E_\gamma$ lies in the range 

\begin{equation}
    \frac{E_\gamma^0}{E_e} \leq E_\gamma \leq \frac{E_e \Gamma_E}{1 + \Gamma_E}.
\end{equation}


At high energies, inverse-Compton emission is sharply concentrated near the electron energy $E_e$: a single scattering typically transfers nearly all of the $e^\pm$ energy to the upscattered photon (as seen in Fig.\,9 of Ref.\,\cite{Esmaili:2015xpa}).

\begin{figure*}
    \centering
    \includegraphics[width=1.0\linewidth]{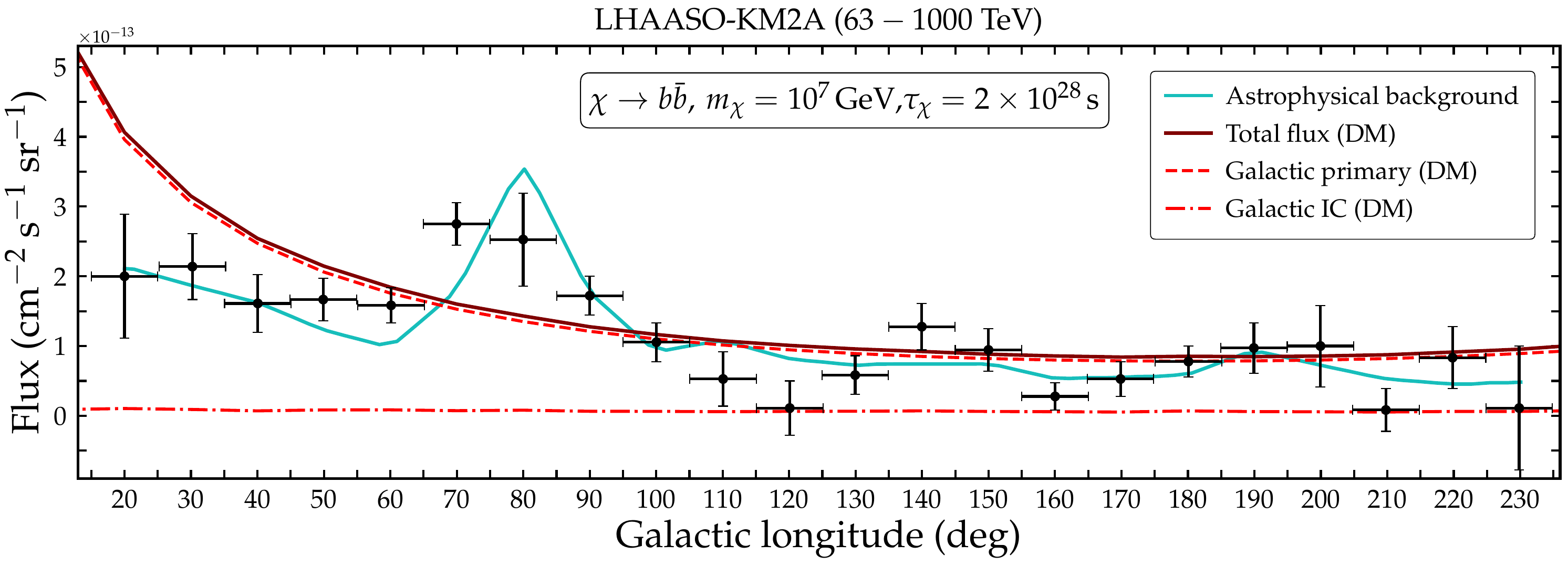}
    \caption{Energy integrated gamma-ray flux as a function of photon energies for the DM decay channel $\chi \to b\bar{b}$ with $m_{\chi} = 10^{7}\,\mathrm{GeV}$ and $\tau_{\chi} = 2 \times 10^{28}\,\mathrm{s}$, along with the LHAASO-KM2A galactic diffuse gamma-ray longitude profile for the energy range 63-1000 TeV. The DM flux, astrophysical background, and data point colour schemes and linestyles are same as Fig.\,\ref{fig:flux_plots}. Given the subdominant contribution and isotropic nature, we do not show the extragalactic DM signal here. The chosen mass and lifetime parameters are excluded by our
analysis. We note that the peak in the astrophysical model at around $80^\circ$ latitude is coming from the Cygnus bubble\,\cite{Chen:2024yin}.} 
    \label{fig:angular}
\end{figure*}

The diffusion halo function, $I_{\mathrm{diff}}(E, E_s, s, b, l)$, is derived by solving the diffusion-loss equation for $e^\pm$ propagation in the Galactic medium. This function encapsulates the effects of spatial diffusion and other transport phenomena experienced by the electrons and positrons as they traverse the Galactic halo. However, at the high-energy regime considered in this study, the diffusion halo function $I_{\mathrm{diff}}$ effectively approaches unity, indicating that energy losses dominate over diffusive effects, thereby simplifying the treatment of electron propagation\,\cite{Cirelli:2010xx,Esmaili:2015xpa}.

\subsection{Extragalactic Components}
 The prompt extragalactic flux is
\begin{equation}
    \frac{d\phi^{\text{EG}}_{\text{prompt}}}{dE_\gamma d\Omega} = \frac{\Omega_\chi\rho_{\text{c}}}{4\pi m_{\chi}\tau_\chi}\int^\infty_0\frac{dz}{H(z)}e^{-\tau_{\gamma \gamma}\left(E_\gamma, z\right)}\frac{dN_\gamma}{dE_\gamma}(E_\gamma\,(1+z)).
\end{equation}

The extragalactic IC component is:
\begin{equation}
\begin{aligned}
    \frac{d\phi^{\text{EG}}_{\text{IC}}}{dE_\gamma d\Omega} = 
    & \frac{\Omega_\chi\rho_{\text{c}}}{2\pi E_\gamma m_{\chi}\tau_\chi} 
     \int^\infty_0\frac{dz\, e^{-\tau_{\gamma \gamma}\left(E_\gamma, z\right)}}{H(z)(1+z)} \\
    & \times \int^{m_{\chi}/2}_{E_\gamma(1+z)}dE_e 
     \frac{P_{\text{IC}}(E_\gamma(1+z),E_e)}{b^{\rm EG}_{\text{IC}}(E_e)} \\
    & \times \int^{m_{\chi}/2}_{E_e}dE'\frac{dN_e}{dE'}\,,
\end{aligned}
\end{equation}
where $\Omega_\chi = 0.27$ is the cosmic DM density parameter, $\rho_{\text{c}} = 1.15 \times 10^{-6}$ GeV cm$^{-3}$ is the critical density, $H(z)$ is the Hubble parameter at redshift $z$, and $\tau_{\gamma \gamma}\left(E_\gamma, z\right)$ accounts for absorption on CMB and EBL.

The total gamma-ray flux then is the sum of all components:
\begin{equation}
    \frac{d\Phi}{dE_\gamma d\Omega} = \left(\frac{d\phi^G_{\text{prompt}}}{dE_\gamma d\Omega} + \frac{d\phi^G_{\text{IC}}}{dE_\gamma d\Omega}\right) + \left(\frac{d\phi^{\text{EG}}_{\text{prompt}}}{dE_\gamma d\Omega} + \frac{d\phi^{\text{EG}}_{\text{IC}}}{dE_\gamma d\Omega}\right)\,.
\end{equation}
This comprehensive treatment of all emission components allows for accurate constraints on DM decay from gamma-ray observations. 

We do not include the extragalactic electromagnetic cascade in our analysis. Photons injected at cosmological distances with energies above a few TeV are efficiently absorbed on the EBL/CMB through $\gamma-\gamma$ pair production on scales much shorter than a Hubble length. The resulting $e^{\pm}$ then IC scatter on background photons and dump the energy into a nearly universal cascade spectrum that piles up in the GeV to sub-TeV range\,\cite{Coppi:1996ze,Murase:2012xs,Murase:2015gea}. At 10--10$^3$ TeV, where LHAASO-KM2A and Tibet AS$_\gamma$ are sensitive, essentially none of this reprocessed power survives, so its contribution to our dataset is negligible. In our modeling we therefore include the prompt and IC contributions  for both Galactic and extragalactic DM decay and omit the extragalactic cascade.

\section{Data \& Analysis}
\label{dataset}

\subsection{Tibet AS$_\gamma$ \,\cite{Neronov:2021ezg}}
While the diffuse gamma-ray measurement by Tibet AS$_\gamma$ focused on the galactic plane region\,\cite{TibetASgamma:2021tpz}, the gamma-ray sky away from galactic plane remains largely unexplored. Gamma-ray induced showers are muon-poor, whereas cosmic-ray induced showers are muon-rich. Ref.\,\cite{Neronov:2021ezg} used the all-sky cosmic ray measurement by Tibet AS$_\gamma$\,\cite{TibetAS:2019kpg} and identified muon-poor showers by using the corresponding muon-cut. Assuming that all the remaining events after applying muon-cut are gamma-ray induced, the authors put a conservative upper limit on the high-latitude ($|b|>20^{\circ}$) diffuse gamma-ray flux. This upper limit is stronger than the previous upper limits from  KASCADE\,\cite{2017ApJ...848....1A}, CASA-MIA\,\cite{PhysRevLett.79.1805}, GRAPES-3\,\cite{minamino2009upper}, and HEGRA\,\cite{2002APh....17..459H}. 

Our analysis focuses on these upper limits, which probe regions where astrophysical backgrounds are significantly reduced compared to the Galactic plane. The interpretation of these limits in the context of DM decay is especially valuable, as the high-latitude regions offer a cleaner potential signal of DM decay due to reduced cosmic-ray interactions and lower interstellar gas density. 

 A DM decay model is excluded whenever, in any energy bin the gamma-ray signal from DM decay exceeds the upper limit

\begin{eqnarray}
    \Phi_{\mathrm{DM}}(m_{\chi},\tau_{\chi}) \;>\;
\Phi_{\text{UL}}\,\,,
\end{eqnarray}
where $\Phi_{\text{UL}}$ is the upper limit on diffuse gamma rays at $|b|>20^{\circ}$, obtained in Ref.\,\cite{Neronov:2021ezg}.

\subsection{LHAASO-KM2A\,\cite{LHAASO:2023gne}}

   KM2A is an extensive air-shower array within LHAASO\,\cite{LHAASO:2019qtb}. KM2A consists of electromagnetic particle detectors (ED) and underground muon detectors (MD), both of which can be used to discriminate between cosmic-ray-induced and gamma-ray-induced air showers. Besides, LHAASO also has the Water Cherenkov Detector Array (WCDA) which primarily focuses on lower energy gamma rays. LHAASO–KM2A has measured diffuse gamma-ray flux  in two longitude windows along the Galactic plane: an inner region, $15^{\circ}<l<125^{\circ}$ with $|b|<5^{\circ}$, and an outer region, $125^{\circ}<l<235^{\circ}$ with the same latitude cut\,\cite{LHAASO:2023gne}.  Throughout the energy range $10~\mathrm{TeV}\le E\le1~\mathrm{PeV}$, the observed flux in both regions exceeds the canonical astrophysical background model from cosmic-ray interactions with the interstellar medium. The excess is about a factor of two to three.  Such a surplus can be accommodated by unresolved sources, spatially extended pulsar-wind nebulae, TeV halos or a better modeling of the cosmic-ray interactions\,\cite{Chen:2024yin,Zhang:2023ajh,He:2025oys,Ambrosone:2025wxc,Castro:2025wgf,Kato:2025gva,DeLaTorreLuque:2025zsv, He:2025oys, Prevotat:2025ktr,Vecchiotti:2024kkz,Kaci:2024lwx,Kaci:2024wra,Dekker:2023six,Shao:2023aoi}.

Previously, LHAASO measurements of diffuse gamma-ray away from the Galactic plane regions and dwarf galaxies have been used to put stringent limits on decaying/annihilating particle DM\,\cite{LHAASO:2022yxw,LHAASO:2024upb}. In this work, we use the measurements presented in Ref.\,\cite{LHAASO:2023gne} to derive limits on heavy decaying DM. We note that the region of interest for Ref.\,\cite{LHAASO:2022yxw} and Ref.\,\cite{LHAASO:2023gne} are different.

To obtain our limits, we assume that the observations made by LHAASO\,\cite{LHAASO:2023gne} are consistent with the expected diffuse gamma-ray model predictions. We use the astrophysical model derived in Ref.\,\cite{Chen:2024yin} as our background model for diffuse gamma-ray. This background model incorporates improved signal leakage models for known sources to explain the discrepancy in the LHAASO measurement. LHAASO collaboration has also used an astrophysical model considering the local cosmic ray spectra and gas column density\,\cite{LHAASO:2023gne}. Given that their measurements exceed their background model, they multiply their model by 2 and 3 for the outer and inner region datasets, respectively. We refer to this as the LHAASO `naive' background model. To compare the dependence of our limits on the choice of the background models, we also use this `naive' background model given in Ref.\,\cite{LHAASO:2023gne} to obtain the corresponding limits on heavy decaying DM. We note that in their recent LHAASO-WCDA analysis\,\cite{LHAASO:2024lnz}, the collaboration has updated their LHAASO-KM2A measurements. These updated measurements are in agreement with the previous dataset within the error bars. In our work, we use the dataset presented in Ref.\,\cite{LHAASO:2023gne}.

Besides the spectral dataset, LHAASO has also measured the energy integrated angular profile for their diffuse gamma-ray flux measurements\,\cite{LHAASO:2023gne}. This can be explained by the astrophysical background model presented in Ref.\,\cite{Chen:2024yin}. Angular flux measurement can better discriminate between astrophysical and DM originated photon flux, due to the inherent difference between the baryon and DM density profiles. We use these angular measurements presented in Ref.\,\cite{LHAASO:2023gne} along with the best-fit astrophysical background model presented in Ref.\,\cite{Chen:2024yin} to obtain limits on heavy DM decay.

A possible contribution from DM decay is tested by 
using a $\chi^{2}$ analysis\,\cite{Cowan:2010js, Lamperstorfer:2015cfg} 

\begin{figure}[!htbp]
\includegraphics[width=\columnwidth]{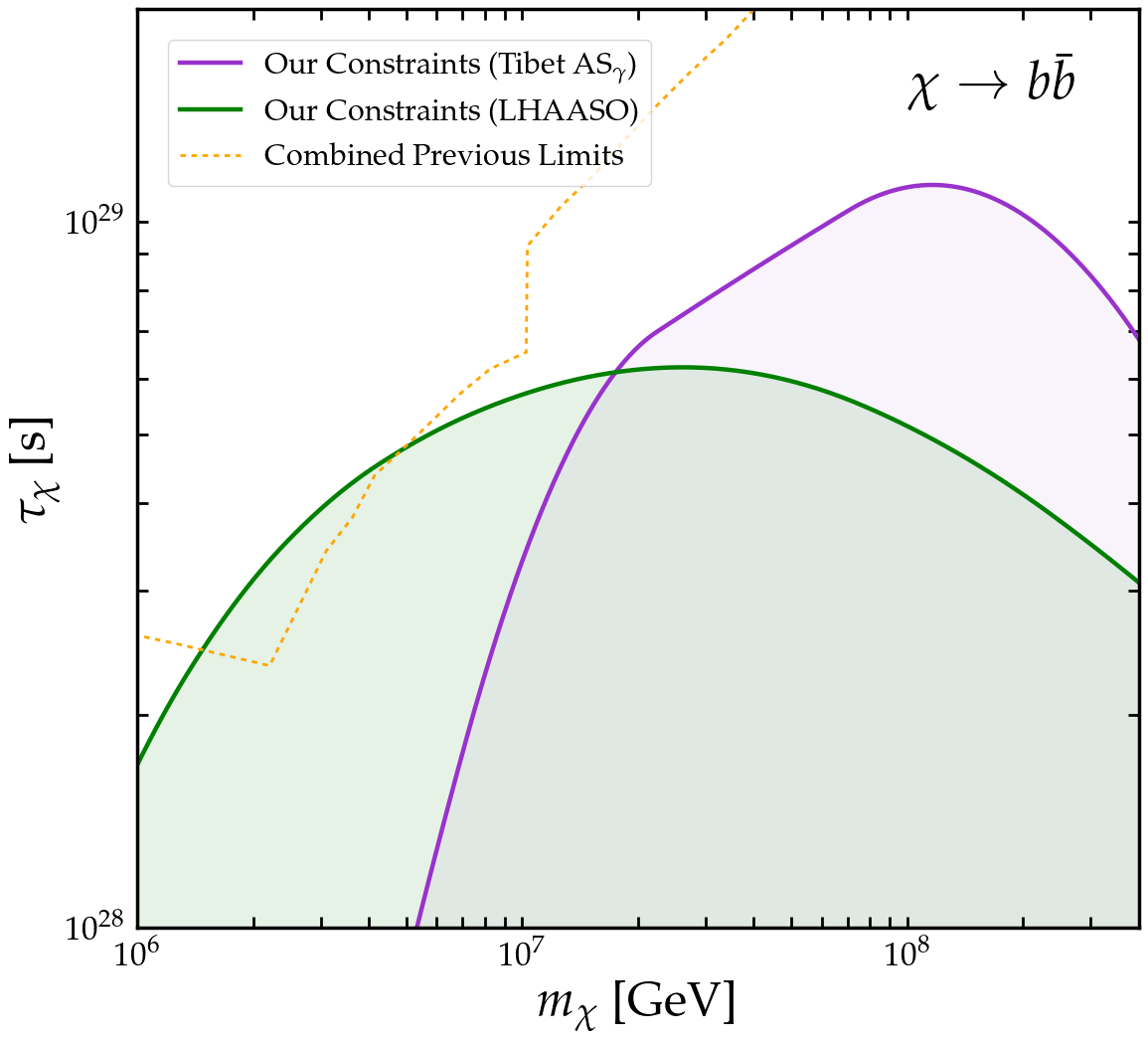}
\caption{Upper limits on DM lifetime for the decay channel $\chi \to b\bar{b}$. Our constraints from Tibet AS$_\gamma$ (Neronov \textit{et al.})\,\cite{Neronov:2021ezg} and LHAASO-KM2A\,\cite{LHAASO:2023gne} datasets are shown by the purple solid and green solid lines, respectively. Previous combined best limit in the parameter space is taken from Refs.\,\cite{Esmaili:2014rma,Cohen:2016uyg,Blanco:2018esa,Aloisio:2025nts,Ishiwata:2019aet,Chianese:2019kyl,LHAASO:2022yxw} (orange dashed line). 
}
\label{fig: bbbar_constraints}
\end{figure}

\begin{eqnarray}
    \chi^{2}(m_{\chi},\tau_{\chi}) \;=\;
      \sum_{i}
      \frac{\bigl[\Phi_{\mathrm{obs}}^i
                 -\Phi_{\mathrm{bkg}}^i
                 -\Phi_{\mathrm{DM}}^i(m_{\chi},\tau_{\chi})\bigr]^{2}}
           {\sigma_i^{2}}\,\,.\nonumber
\end{eqnarray}
In the above equation, $\Phi_{\mathrm{obs}}^i$ and  $\sigma^i$ are the measured flux and the flux error in the $i^{\rm th}$ energy bin, respectively. The gamma-ray flux from astrophysical background model and DM decay at the $i^{\rm th}$ energy bin, are given by $\Phi_{\mathrm{bkg}}^i$ and $\Phi_{\mathrm{DM}}^i(m_{\chi},\tau_{\chi})$, respectively.
The combined statistical and systematic uncertainties (added in quadrature) quoted in Ref.\,\cite{LHAASO:2023gne}\,(table S2, S3) are used as $\sigma$ in the above equation. We calculate the $95\%$ CL constraints on the DM lifetime from the following equation.
\begin{eqnarray}
    \chi^{2}(m_{\chi},\tau_{\chi})-\chi^{2}_{\min}=2.71\,,
\end{eqnarray}
where $\chi^{2}_{\min}$ corresponds to the minimum value of $\chi^{2}$ for a particular DM mass.

Unlike LHAASO-KM2A, the Tibet AS$_\gamma$ constraints are derived by requiring that the DM signal does not 
exceed the flux upper limits of Ref.\cite{Neronov:2021ezg} in any energy bin. Since these flux upper 
limits assume all surviving muon-poor events are gamma-ray induced with no astrophysical 
background subtracted, our derived bounds are conservative. A proper 
background-subtracted analysis at 95\% CL would yield bounds that are stronger than those presented here.

\begin{figure*}
	\begin{center}
		\includegraphics[width=\columnwidth]{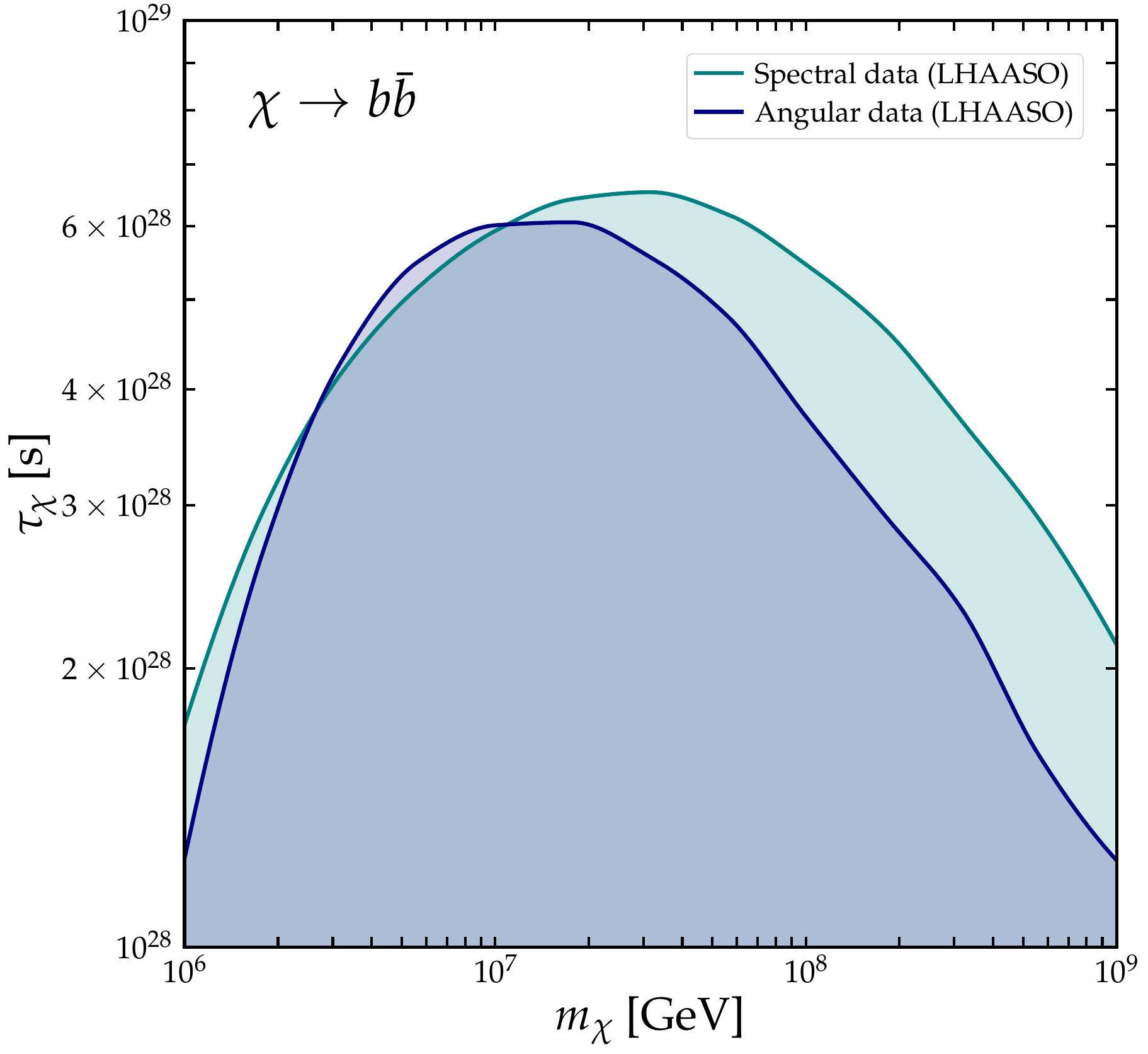}~~
		\includegraphics[width=\columnwidth]{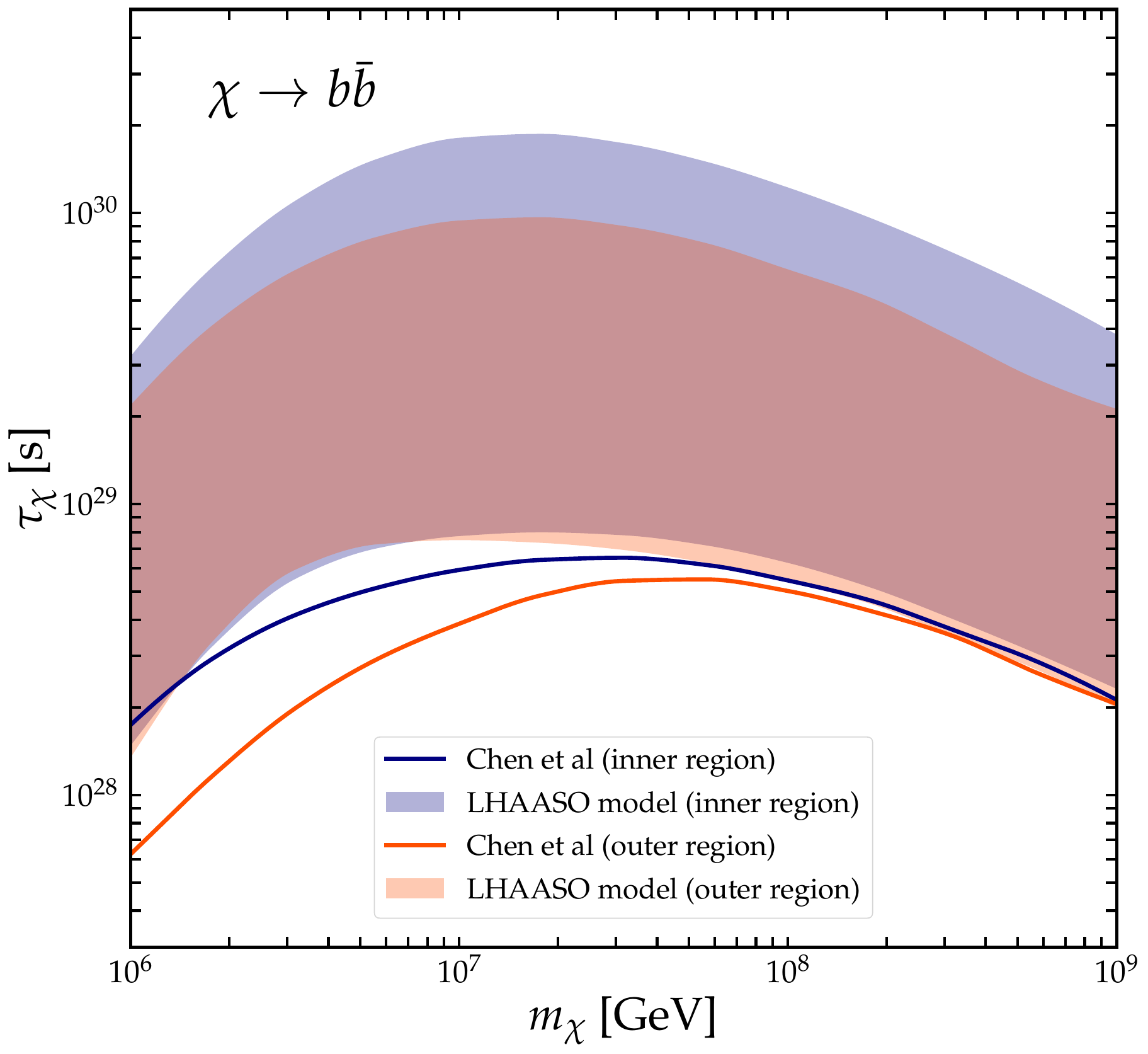}~~\\	
		\caption{(\textbf{Left panel}) Comparison of our bounds from analyzing the spectral datasets (teal solid line) and angular datasets (blue solid line) from LHAASO-KM2A measurement\,\cite{LHAASO:2023gne}. For spectral data, we use our benchmark limit from Fig.\,\ref{fig: bbbar_constraints}. For the angular dataset, we use a longitudinal profile in the energy range 63-1000 TeV, which provides the strongest bound. For both these limits, we have assumed the astrophysical background model presented in Ref.\,\cite{Chen:2024yin}. (\textbf{Right panel}) Dependence of our LHAASO-KM2A limits on the choice of astrophysical background model. We show the results for our benchmark background model from Ref.\,\cite{Chen:2024yin} by solid blue (inner region) and solid maroon (outer region) lines. For comparison, we show the limits using the `naive' background model used by LHAASO\,\cite{LHAASO:2023gne}. The limits using the inner (denoted as `$\times 3$' in Ref.\,\cite{LHAASO:2023gne}) and outer (denoted as `$\times 2$' in Ref.\,\cite{LHAASO:2023gne}) region background models are shown by the blue shaded and orange shaded regions, respectively. The shaded regions show the uncertainty in the background model shown in Ref.\,\cite{LHAASO:2023gne}. Note that the $y$ axis range in the two panels is not the same.}
		\label{fig:AngularLimit}
	\end{center}	
\end{figure*}

In our analysis, we do not take into account the finite energy resolution of the high-energy gamma-ray telescopes. For decay channels considered here (except for the two photon decay channel), the resulting photon spectra is smooth in nature. As a result the energy resolution of the instrument has a subdominant effect. Additionally, the systematic uncertainty arising from the choice of the diffuse astrophysical background model is much larger than effects of energy resolution (which is 20\% at 100 TeV for LHAASO-KM2A)\,\cite{LHAASO:2023gne}.  We find that the uncertainty in our limits due to the choice of astrophysical background model can be as large as factor of $\sim$ 10 (see right panel  of Fig.\,\ref{fig:AngularLimit}).

\section{Results}
\label{results}

In Fig.\,\ref{fig:flux_plots} we show the differential gamma-ray flux expected from heavy DM decay ($\chi \to b\bar{b}$) with $m_\chi = 10^7~\mathrm{GeV}$ and $\tau_\chi = 2\times 10^{28}~\mathrm{s}$. The DM contributions are separated into `Galactic primary' (red dashed line), `Galactic IC' (red dot-dashed line), and `extragalactic' (red dotted line) components. The total DM flux is shown by the solid maroon line. At the highest energies, approaching the PeV scale, the spectrum is dominated by the Galactic prompt photons directly produced from hadronic decays. At lower energies ($E_\gamma \lesssim 10^4~\mathrm{GeV}$), the secondary IC component arising from upscattering of ambient CMB, SL, and IR photons by DM-induced $e^\pm$ becomes increasingly important and even exceeds the prompt flux in some energy range \cite{Leung:2023gwp}. The extragalactic component, including both prompt and IC emission, is heavily attenuated by the CMB/EBL through $\gamma\gamma$ absorption, and therefore contributes negligibly at $E_\gamma \gtrsim 10~\mathrm{TeV}$. 

In Fig.\,\ref{fig:flux_plots} we compare the DM signal with the diffuse gamma-ray upper limits from Tibet AS$_\gamma$\,\cite{Neronov:2021ezg} (green upper limits) and the diffuse gamma-ray flux measurements from LHAASO-KM2A\,\cite{LHAASO:2023gne} (black data points) in the left and right panels, respectively. The LHAASO-KM2A flux measurement error bars include both systematic and statistical errors, added in quadrature. For the LHAASO measurements, the total DM-induced flux is plotted along with the astrophysical background model from Ref.\,\cite{Chen:2024yin}. For the benchmark DM parameters considered, the DM-induced flux exceeds the diffuse gamma-ray upper limits from Tibet AS$_\gamma$. Similarly, for the  LHAASO diffuse measurements, the  DM signal and the astrophysical background together overshoot the measurements across a substantial part of the energy range. This benchmark DM lifetime is therefore excluded by our analysis.

In Fig.\,\ref{fig:angular} we show the energy integrated longitude profile of the DM signal (for benchmark values used in Fig.\,\ref{fig:flux_plots}), along with the LHAASO measurements\,\cite{LHAASO:2023gne} and astrophysical model from Ref.\,\cite{Chen:2024yin}. The color scheme and linestyles are same as Fig.\,\ref{fig:flux_plots}. Evidently, the benchmark decay value of $\tau_\chi = 2\times 10^{28}\,\mathrm{s}$ is also excluded for the angular data.

The 95\% C.L. upper limits on DM lifetime for $\chi \to e^+e^-$ and $\chi \to b\bar{b}$ are shown in Figs.\,\ref{fig:ee_constraints} and \ref{fig: bbbar_constraints}, respectively using the Tibet AS$_\gamma$ (purple solid line) and LHAASO (green solid line) measurements. The combined previous limits are shown by the orange dashed line. Leptonic final states are expected to produce harder photon spectra, whereas hadronic final states produce softer photon spectra. This effect can be seen in our limits presented in Figs.\,\ref{fig:ee_constraints} and \ref{fig: bbbar_constraints}. For both Tibet AS$_\gamma$ and LHAASO datasets, the $\chi \to e^+e^-$ limits are peaked at lower DM masses compared to $\chi \to b\bar{b}$. The impact of IC emission is particularly pronounced for the leptonic decay channels, such as $\chi \to e^+e^-$, where secondary gamma rays from electron/positron upscattering of background photons contribute substantially to the total flux.



We have considered three background photon fields (CMB, SL, and IR) in calculating the IC contribution. The CMB component contributes most significantly due to its homogeneous nature and well-known properties. The impact of SL and IR backgrounds varies with position in the galaxy and becomes less significant at high latitudes\,\cite{Vernetto:2016alq}.

The extragalactic contribution, both from prompt and IC emission, has been included in our analysis. While subdominant compared to galactic components, it provides a more precise analysis for DM decay.

As mentioned before, we also perform our analysis for the angular diffuse gamma-ray measurements presented in Ref.\,\cite{LHAASO:2023gne}. The comparison of the results for the $\chi \to b\bar{b}$ channel is presented in Fig.\,\ref{fig:AngularLimit} (left panel). Here the angular analysis bound includes the combined best limits from LHAASO longitudinal profiles in the energy ranges $10-63$ TeV and $63-1000$ TeV\,\cite{LHAASO:2023gne}.  The results from the spectral analysis (teal solid line) is dominant over the angular analysis (blue solid line) in most parts of the parameter space. This can be attributed to the poor fit of the best-fit astrophysical background model for the angular datasets, as shown in Ref.\,\cite{Chen:2024yin}. The limits using the longitude profile show improvement around $\sim 5\times10^6$ GeV DM mass for $\chi \to b\bar{b}$ channel. But given the improvement is not significant, we choose to use the spectral analysis bounds over the angular analysis bounds for all the limits presented in this work.

Our LHAASO-KM2A limits do depend on the choice of the diffuse gamma-ray background model. In Fig.\,\ref{fig:AngularLimit} (right panel), we show the dependence of our bounds on various choices of the astrophysical background models. The blue and orange curves show the results with the benchmark background model used in our work (taken from  Ref.\,\cite{Chen:2024yin}). For comparison, we show our limits with the `naive' background model presented in Ref.\,\cite{LHAASO:2023gne}, (blue and orange shaded regions). The limits grow stronger or weaker depending on how well the background model agrees with the flux measurements.  Our limits can vary at most by a factor of $\sim 5$ depending on the choice of the background model. As evident from Fig.\,\ref{fig:AngularLimit} (right panel), our benchmark choice of background model yields the most conservative bounds on decaying DM. Besides, we note that for LHAASO-KM2A, the inner region datasets provide stronger limits than the outer region, given the enhanced DM density towards the inner region.

We have also repeated our analysis with other possible DM density profiles, like the Einasto and Isothermal profiles. Our limits change at most by $\sim 2$\%. This is due to the fact that the datasets used in this work are all observations sufficiently away from the Galactic centre where the dependence on the DM density profile is prominent.

Besides $\chi \to e^+e^-$ and $\chi \to b\bar{b}$, we have also evaluated the limits for various other SM final states,  $\chi \to\{u\bar{u},\,d\bar{d},\,s\bar{s},\,c\bar{c},\,t\bar{t},\,gg,\,ZZ,\,W^+W^-,\,hh,\,\tau^+ \tau^-,
\gamma\,\gamma,\,\mu^+ \mu^-, \\ 
\nu_e \bar{\nu}_e, \,\nu_\mu \bar{\nu}_\mu,\,\nu_\tau \bar{\nu}_\tau\}$. For all these final states, our limits are stronger than the previous limits in some parts of the parameter space.



\section{Discussion and Conclusion}
\label{discussion}
In this work, using the recent gamma-ray measurements from the Tibet AS$_\gamma$ and LHAASO experiments, we have derived stringent constraints on heavy DM decaying to various SM final states. We consider both Galactic and extragalactic DM contributions in the total signal, systematically taking into account the prompt and IC contributions. For deriving our limits, we have assumed various diffuse gamma-ray background models that can explain the recent LHAASO measurements. Given the substantial dependence of our LHAASO-KM2A limits on the background model, we have shown the most conservative limits.  Our limits are also robust against the variation of DM density profiles. In future, with a better understanding of the diffuse gamma-ray sky, including the various unresolved sources, these limits can be improved.

For simplicity, the limits presented in this work assume 100\% branching ratio for a particular DM decay channel. For a realistic DM model, the branching ratio to various SM channels may change. In that case our bounds can be scaled accordingly to obtain the corresponding limits.

Recently, using WCDA, the LHAASO collaboration has also made observations of diffuse gamma rays in the energy range of 1-25 TeV,\cite{LHAASO:2024lnz}. This measurement bridges the gap between lower energy gamma-ray measurements by Fermi-LAT\,\cite{Zhang:2023ajh} and higher energy measurements by LHAASO-KM2A\,\cite{LHAASO:2023gne}. With the procedure outlined in this work, one can in principle use the flux measurements from WCDA and KM2A in Ref.\,\cite{LHAASO:2024lnz} along with a robust astrophysical background model, to put limits on heavy DM decaying to various final states. In these energy ranges, the contributions from cascaded gamma-rays from DM decay will be important\,\cite{Coppi:1996ze,Blanco:2018bbf,Capanema:2024nwe}. We leave this for future analysis.

Similar to the Tibet AS$_\gamma$ experiment, LHAASO has also recently measured all-particle cosmic-ray flux\,\cite{LHAASO:2024knt}. Given the gamma-ray event selection efficiency of LHAASO, one can thus derive an upper limit on the diffuse gamma-ray away from the galactic plane, following the same procedure outlined in Ref.\,\cite{Neronov:2021ezg}. These diffuse gamma-ray upper limits can then be utilized to put limits on heavy decaying DM. Such an analysis can be interesting in the context of heavy DM.

DM annihilation to various SM final states is yet another way of producing high-energy gamma rays that can be constrained by Tibet AS$_\gamma$ and LHAASO measurements. In the standard WIMP scenario, DM annihilation is restricted beyond $\sim 100$ TeV DM mass, also known as the `unitarity limit'\,\cite{Griest:1989wd,Smirnov:2019ngs}. In non-minimal scenarios however, one can violate this limit and have heavier DM whose relic abundance is set by annihilation\,\cite{Berlin:2016vnh,Harigaya:2016nlg, Tak:2022vkb}. Thus, high-energy gamma-ray measurements by Tibet AS$_\gamma$ and LHAASO can provide stringent limits on such heavy annihilating DM. We plan to investigate this in a forthcoming analysis\,\cite{Dubey:2025ab}.

\textit{Note added:} While this work was in progress, we came to know about a work in a similar direction by Boehm \textit{et al.}\,\cite{Boehm:2025qro}.

\section*{Acknowledgements} 

We especially thank Ranjan Laha and Tarak Nath Maity for detailed discussions.
We thank Ranjan Laha for helpful comments on the
manuscript. We also thank Debajit Bose, Marco Chianese, Stefano Morisi, and Kenny C.\,Y.\,Ng for useful comments and discussions.  A.D.
acknowledges the financial support provided by the Ministry of Education (MoE), Government of India.
A.K.S. acknowledges the Ministry of Human Resource Development, Government of India, for financial support via the Prime Minister’s Research Fellowship (PMRF).

\bibliographystyle{JHEP}
\bibliography{ref.bib}

\clearpage
\onecolumngrid
\thispagestyle{plain}

\appendix
\section{Constraints on other SM final states} 
In this section we provide the upper limits on DM lifetime for different SM final states derived in this work. We exclude the channels, $\chi \to e^{+} e^{-}$ and $\chi \to b\bar{b}$, as those limits are already presented in the main text. The previous limits for all these final states are taken from Refs.\,\cite{Esmaili:2014rma,Cohen:2016uyg,Blanco:2018esa,IceCube:2018tkk,Kachelriess:2018rty,Ishiwata:2019aet,Chianese:2019kyl,LHAASO:2022yxw,LHAASO:2024upb}.

\begin{figure}[H]
  \centering
  \adjustbox{max width=\textwidth, max totalheight=0.78\textheight, center}{%
 \begin{tabular}{@{}ccc@{}}
      \includegraphics[width=0.32\textwidth]{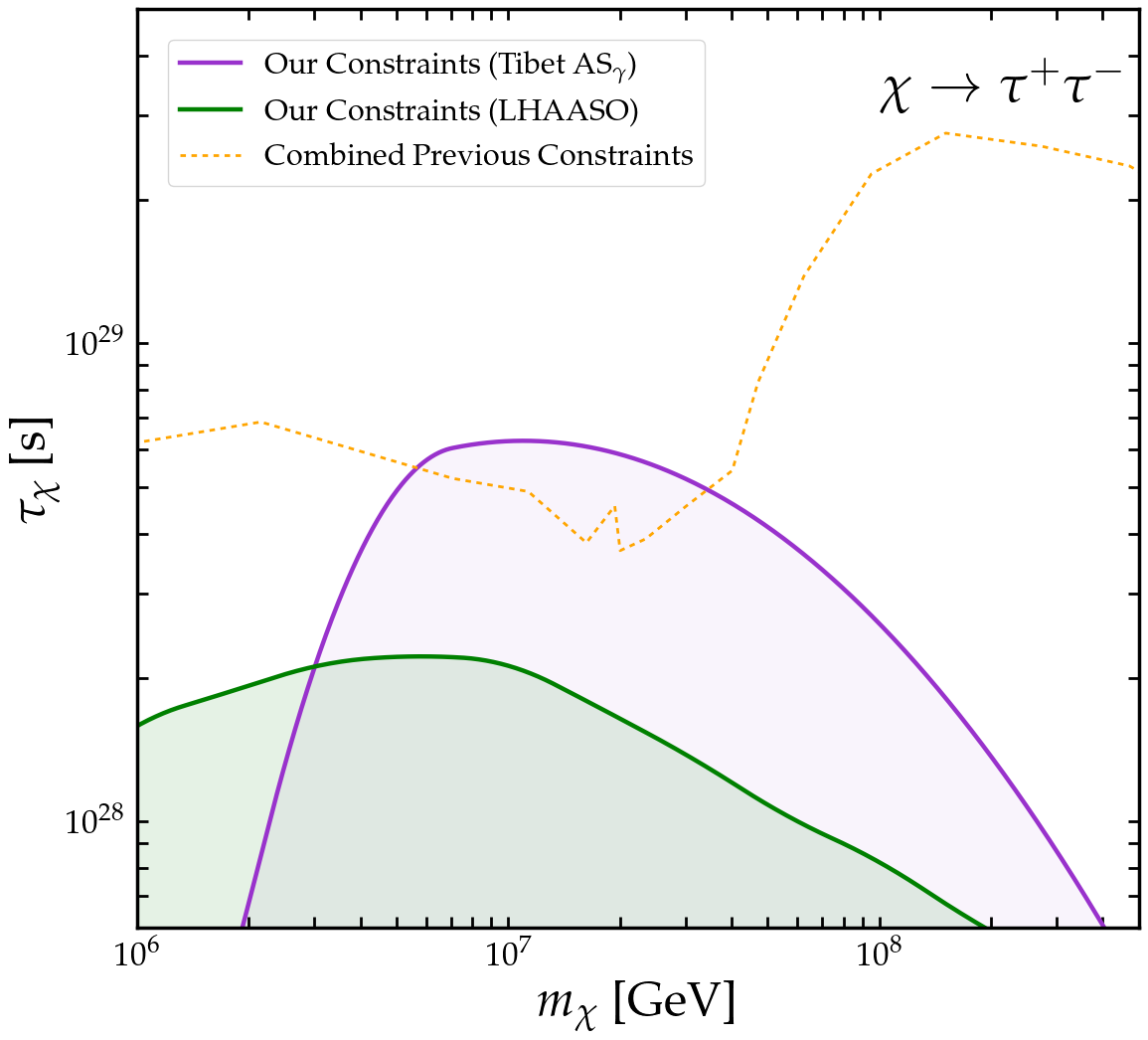} &
      \includegraphics[width=0.32\textwidth]{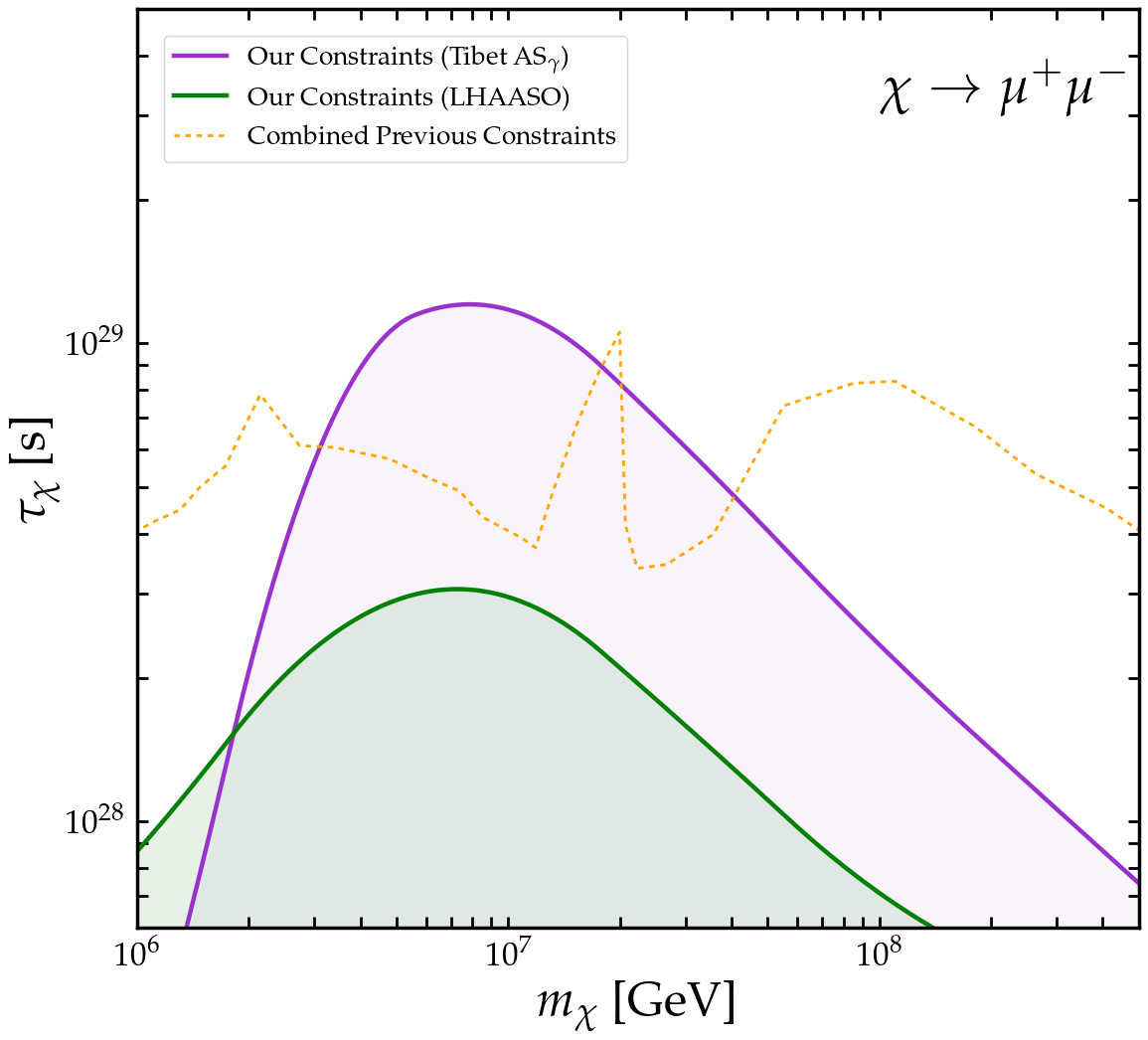} &
      \includegraphics[width=0.32\textwidth]{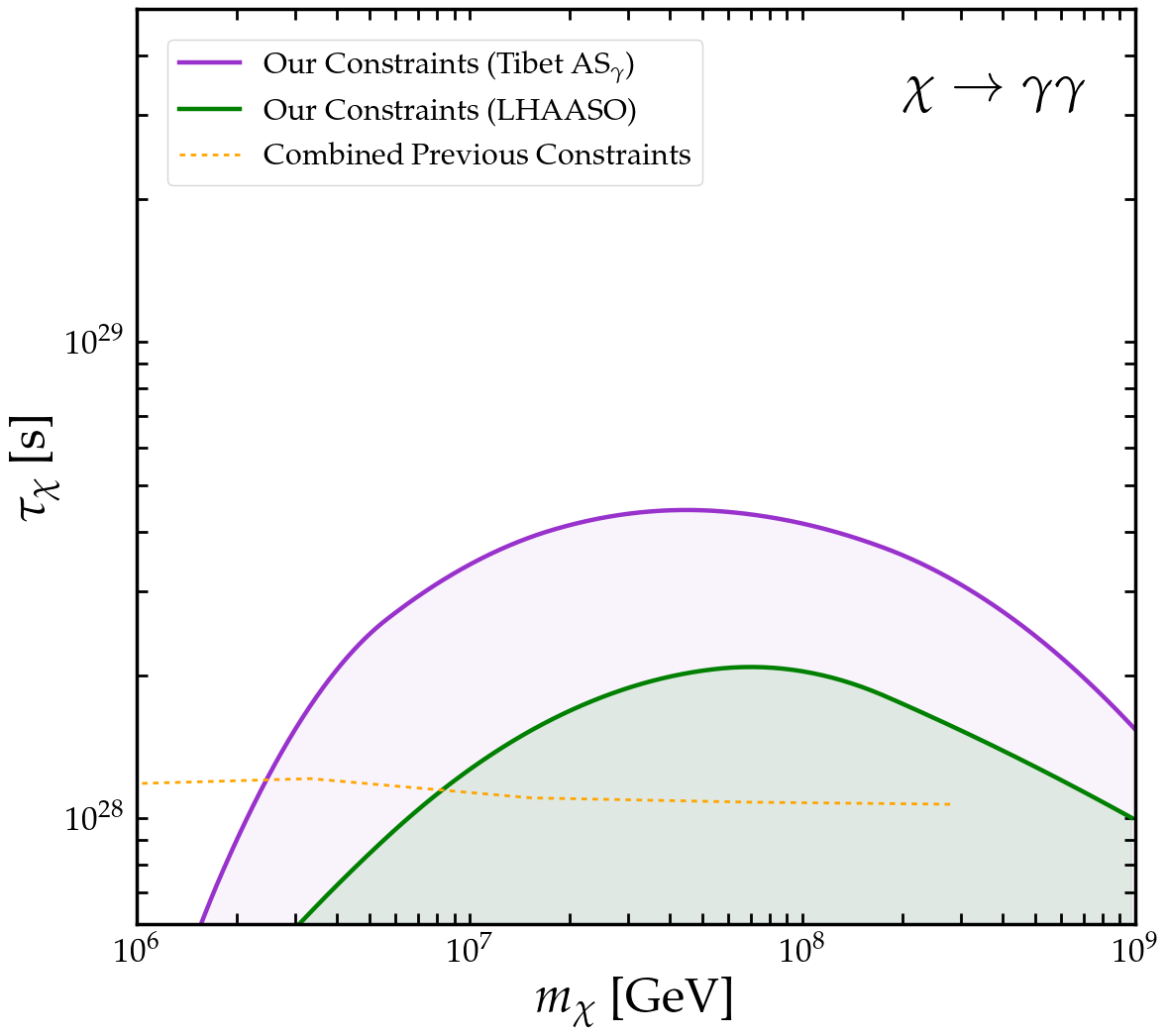} \\
      \includegraphics[width=0.32\textwidth]{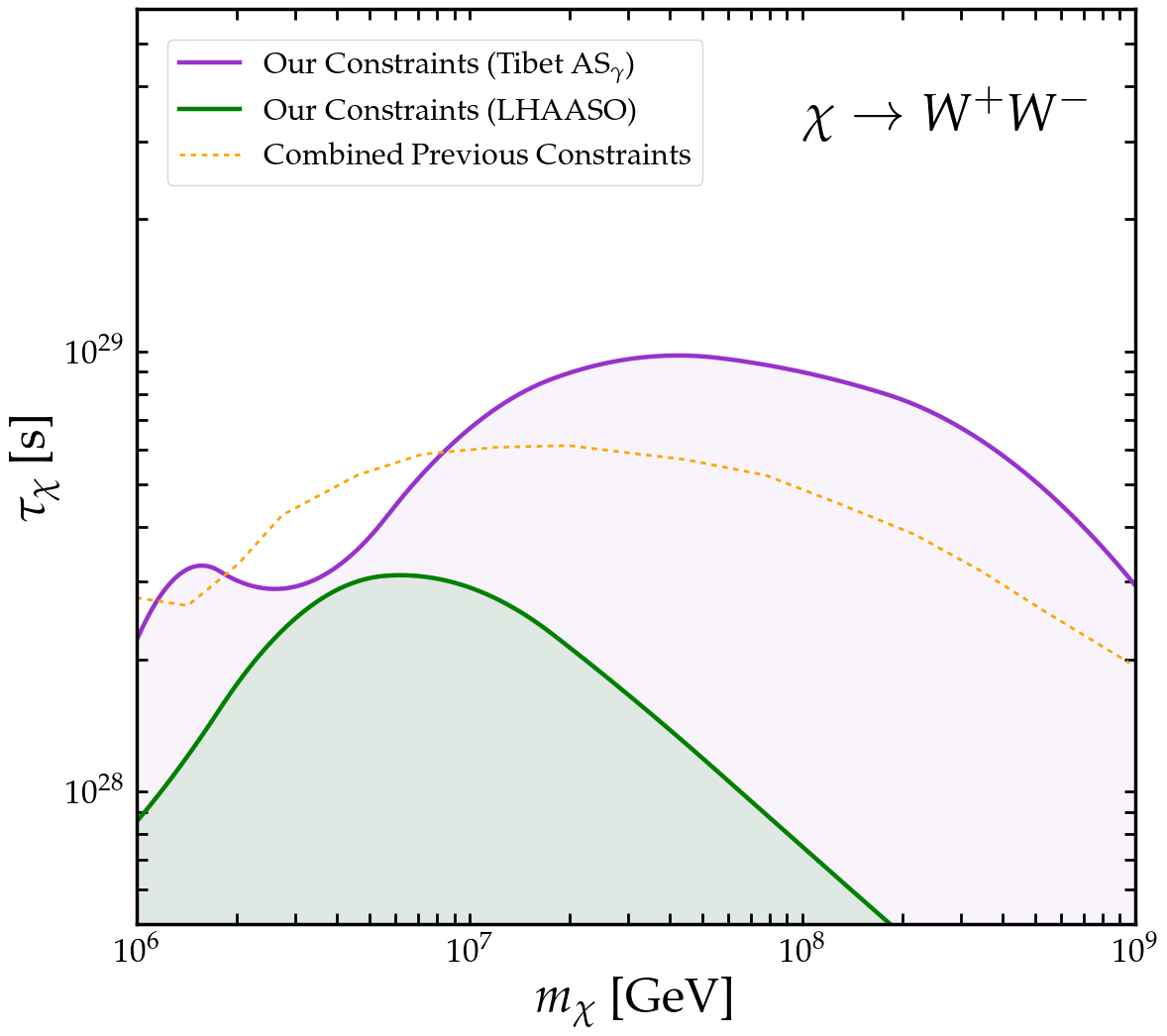} &
      \includegraphics[width=0.32\textwidth]{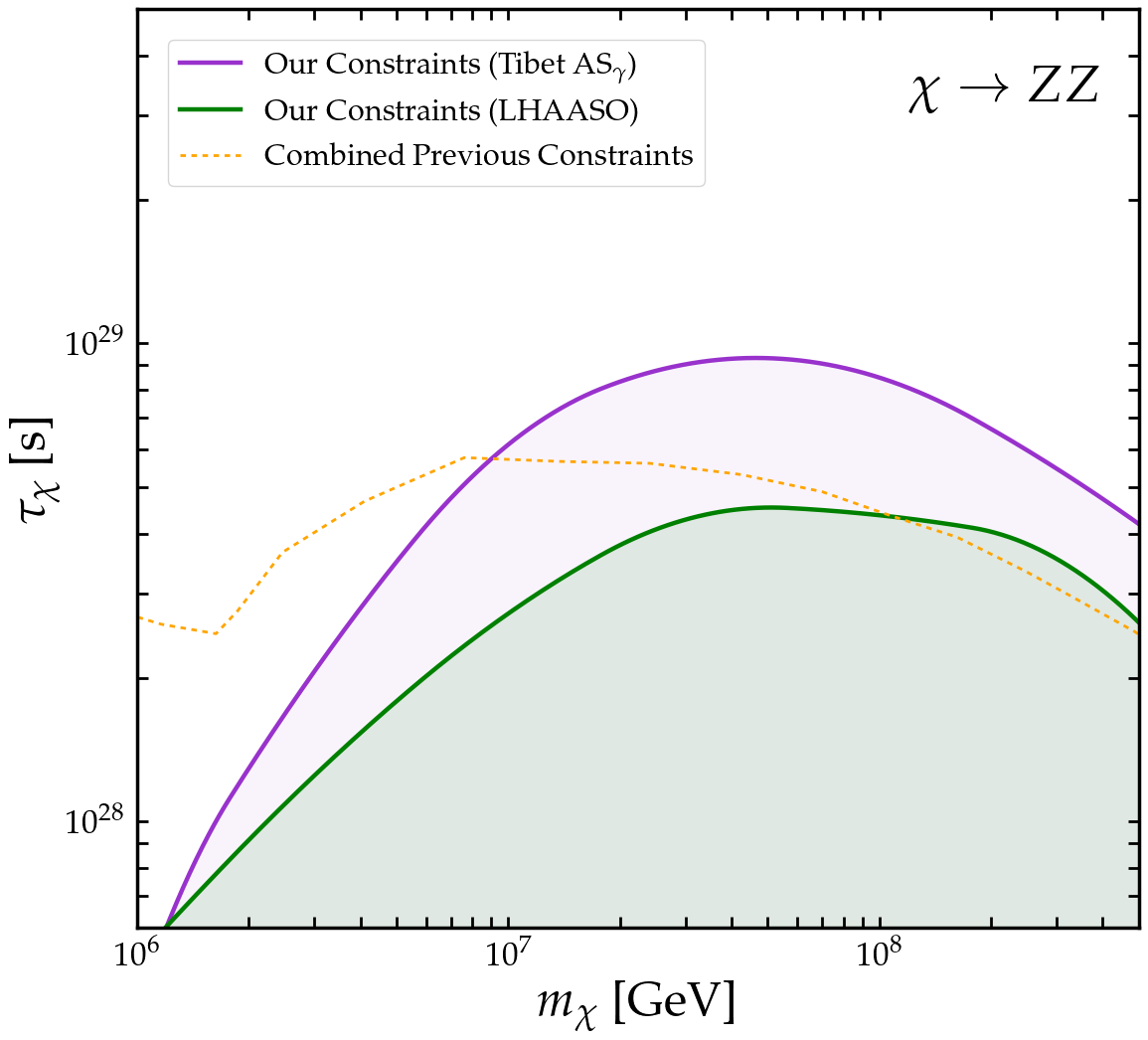} &
      \includegraphics[width=0.32\textwidth]{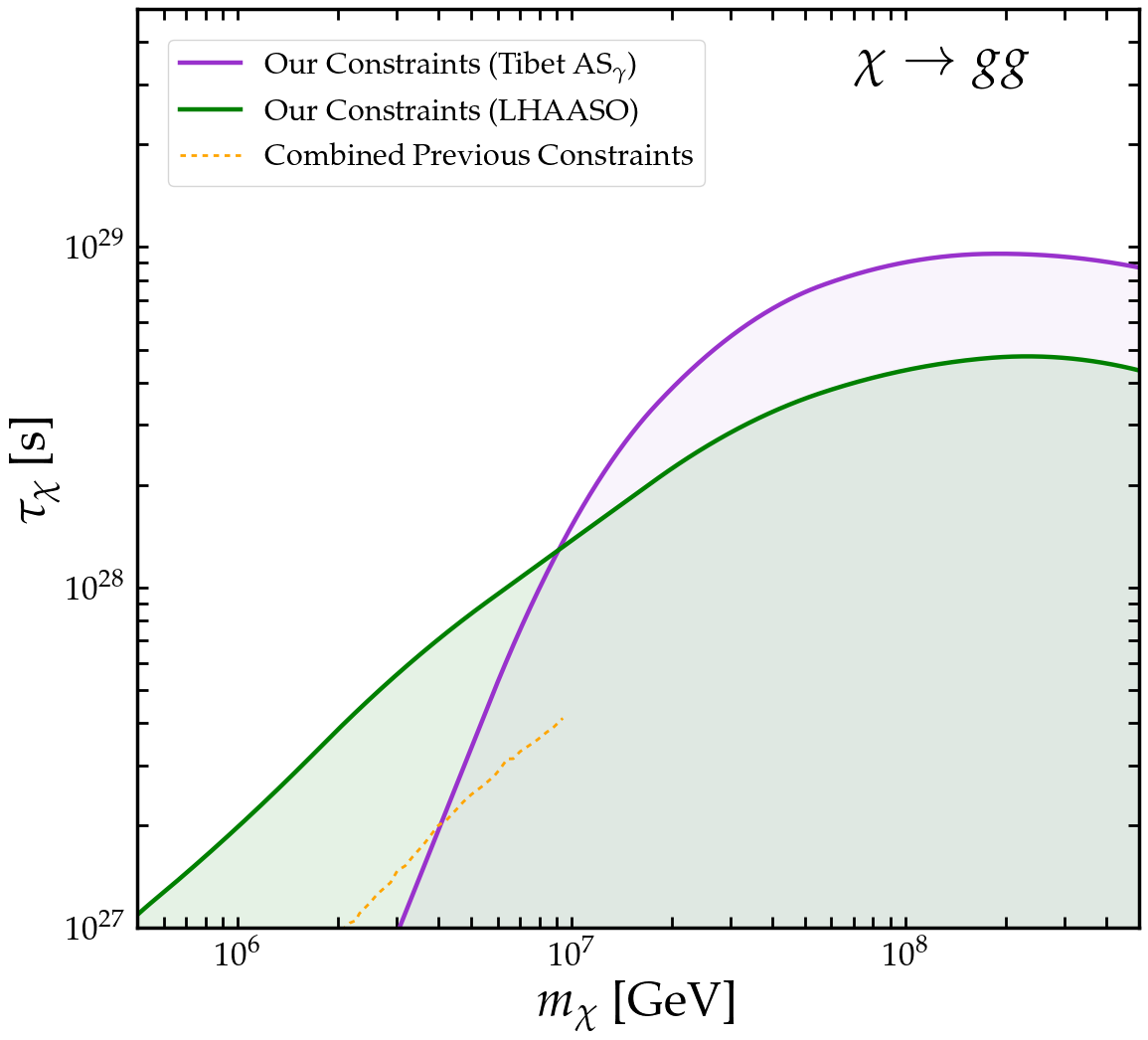} \\
      \includegraphics[width=0.32\textwidth]{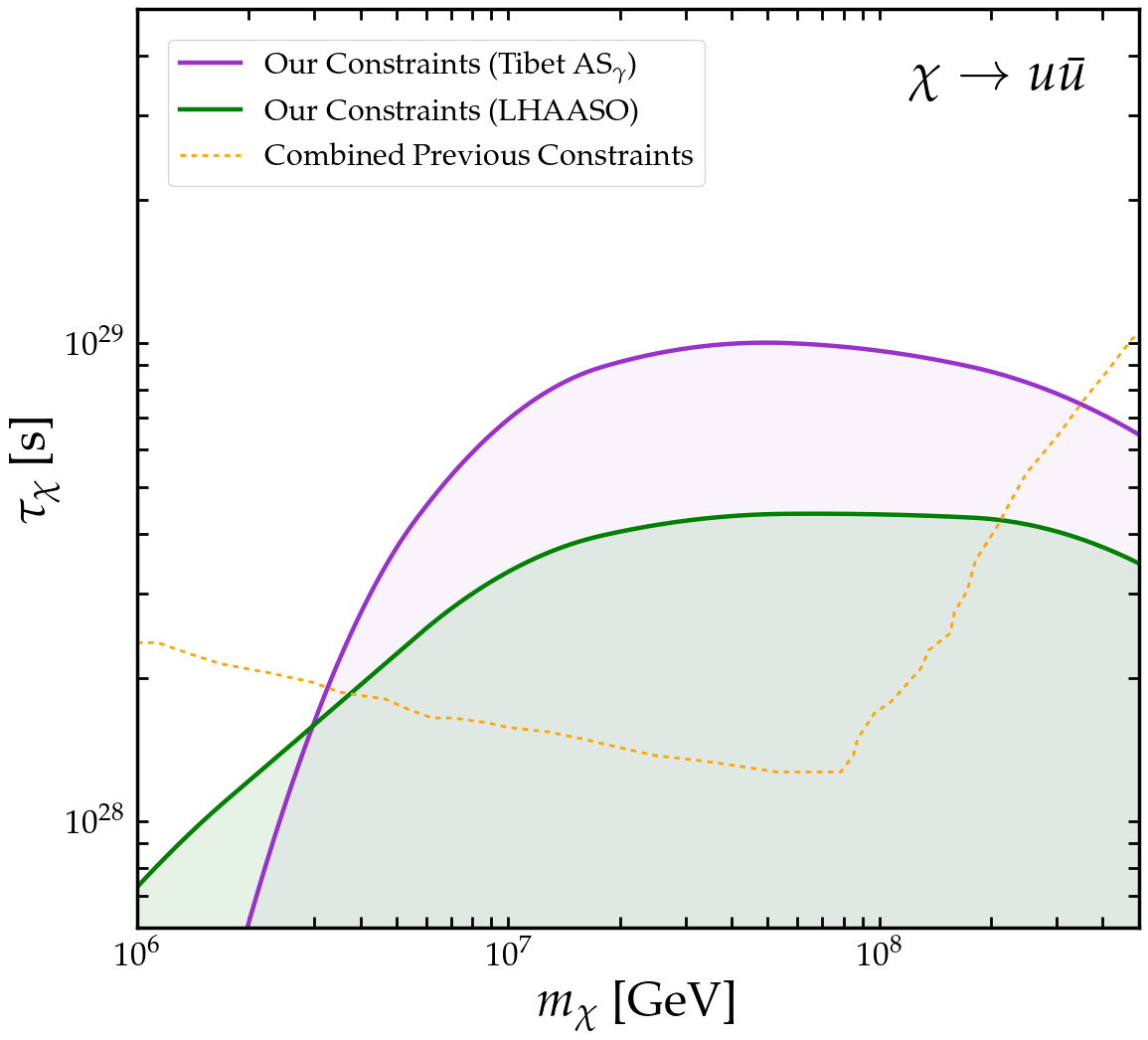} &
      \includegraphics[width=0.32\textwidth]{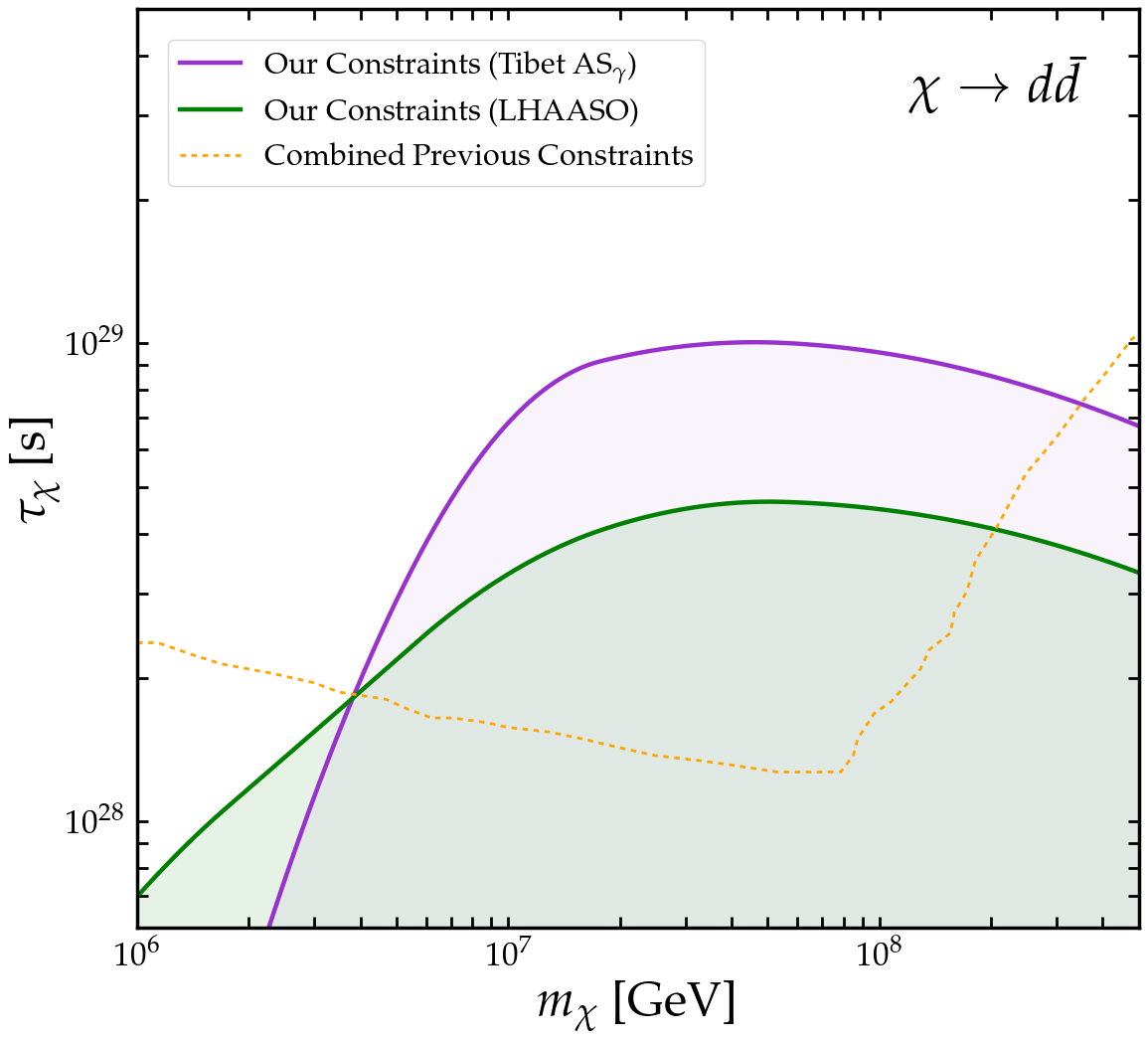} &
      \includegraphics[width=0.32\textwidth]{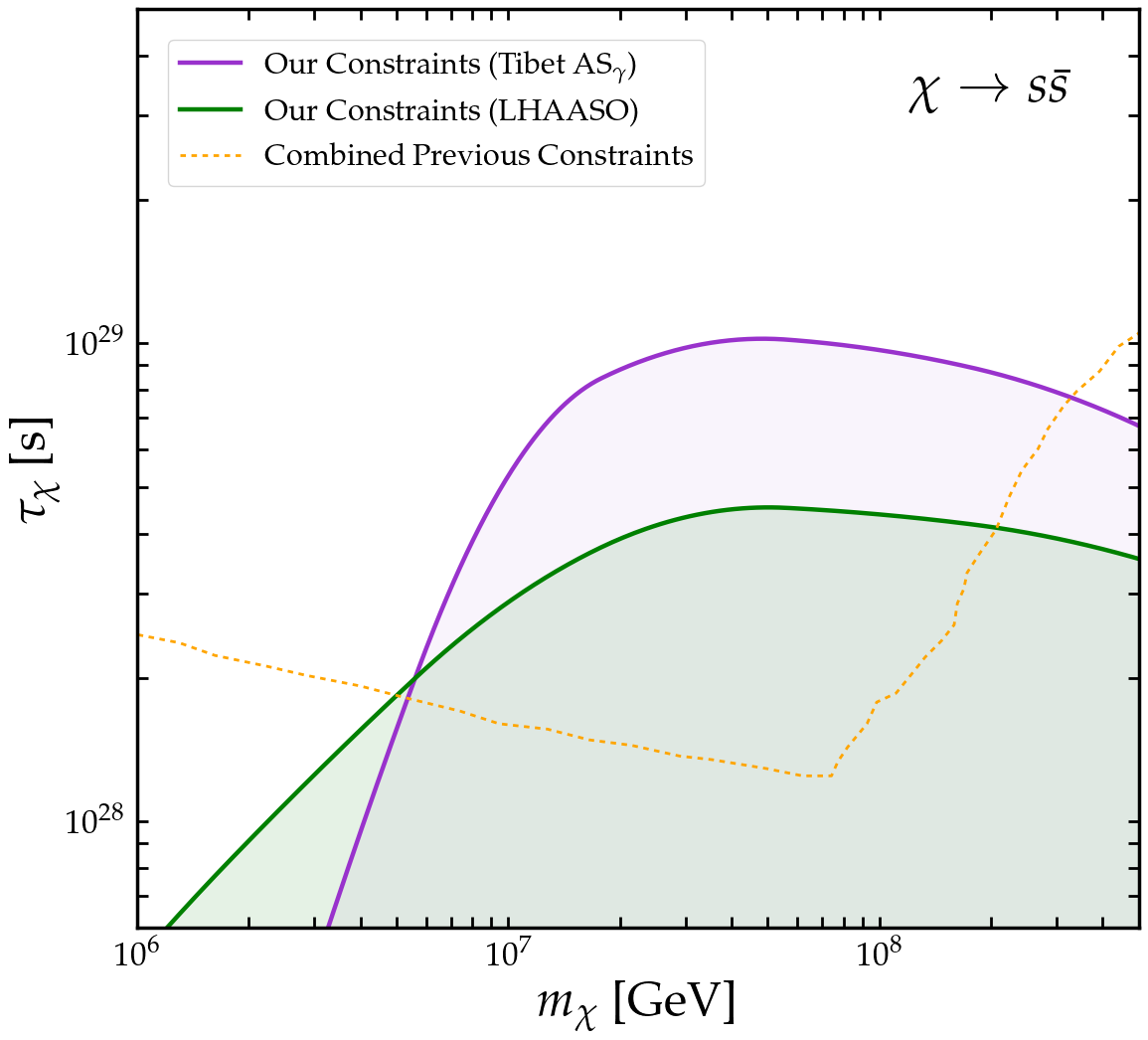} \\
    \end{tabular}%
  }

  \caption{Bounds on DM lifetime for different DM decay channels. Our bounds from Tibet AS$\gamma$ (Neronov \textit{et al.})\,\cite{Neronov:2021ezg} and LHAASO-KM2A\,\cite{LHAASO:2023gne} datasets are shown by the purple solid and green solid lines, respectively. Previous combined best bound in the parameter space is taken from Refs.\,\cite{Esmaili:2014rma,Cohen:2016uyg,Blanco:2018esa,IceCube:2018tkk,Bhattacharya:2019ucd,Ishiwata:2019aet,Chianese:2019kyl,LHAASO:2022yxw,LHAASO:2024upb} (orange dashed line). For ease of comparison, the $y$ axis range is kept the same across all the plots. }
  \label{fig:supp-constraints-1}
\end{figure}

\begin{figure}[H]
  \centering
  \adjustbox{max width=\textwidth, max totalheight=0.78\textheight, center}{%
    \begin{tabular}{@{}ccc@{}}
      \includegraphics[width=0.32\textwidth]{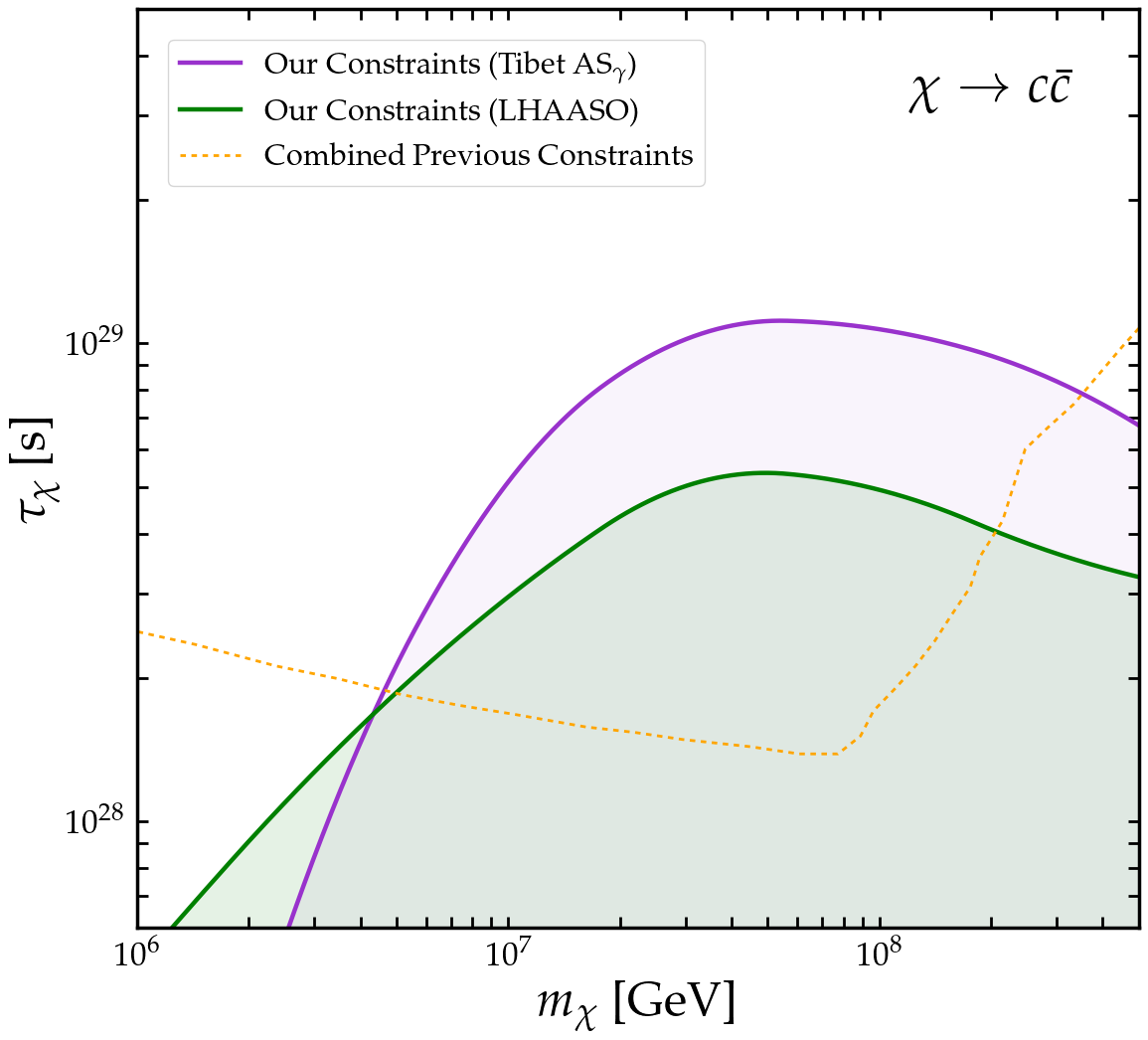} &
      \includegraphics[width=0.32\textwidth]{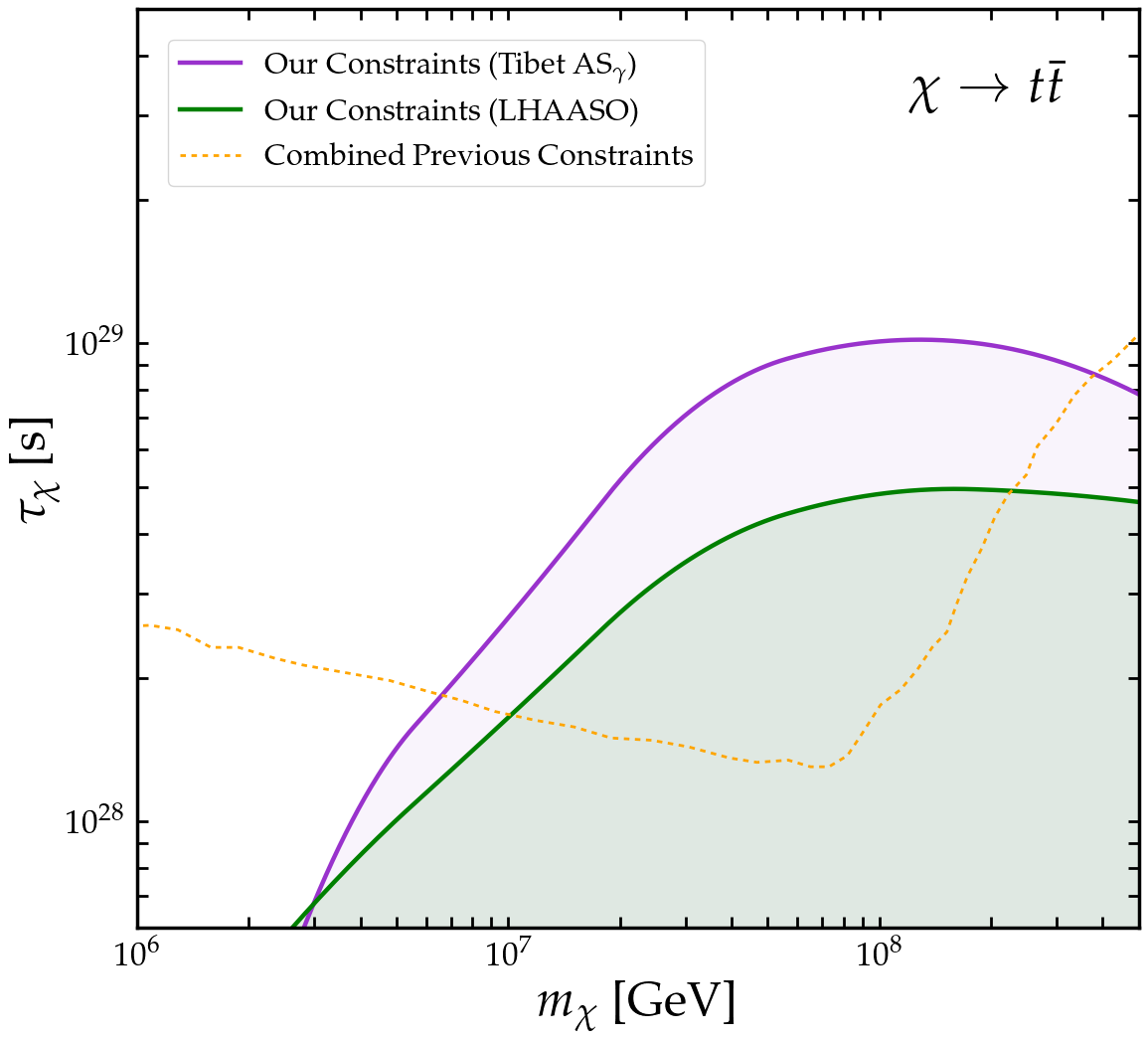} &
      \includegraphics[width=0.32\textwidth]{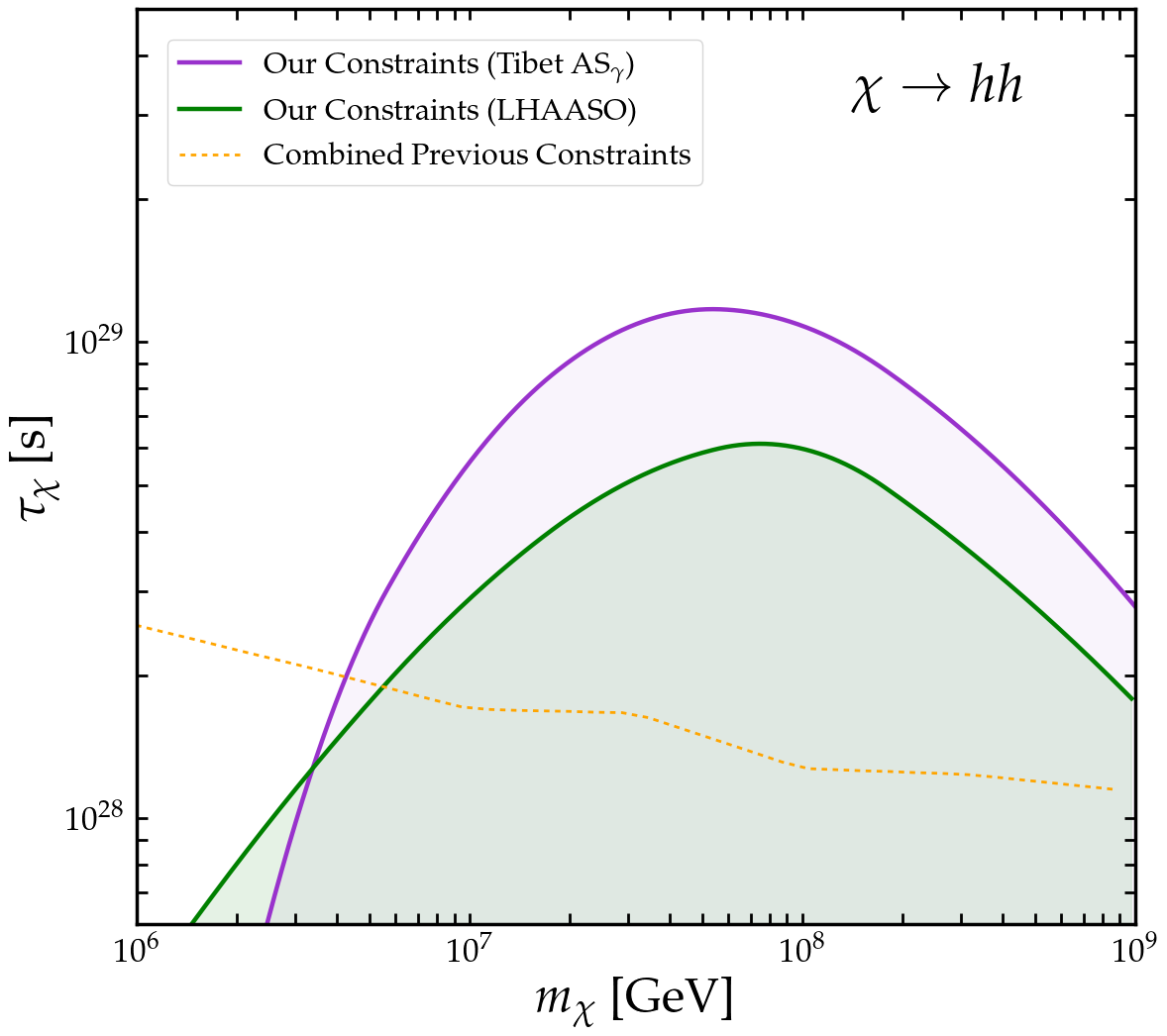} \\
      \includegraphics[width=0.32\textwidth]{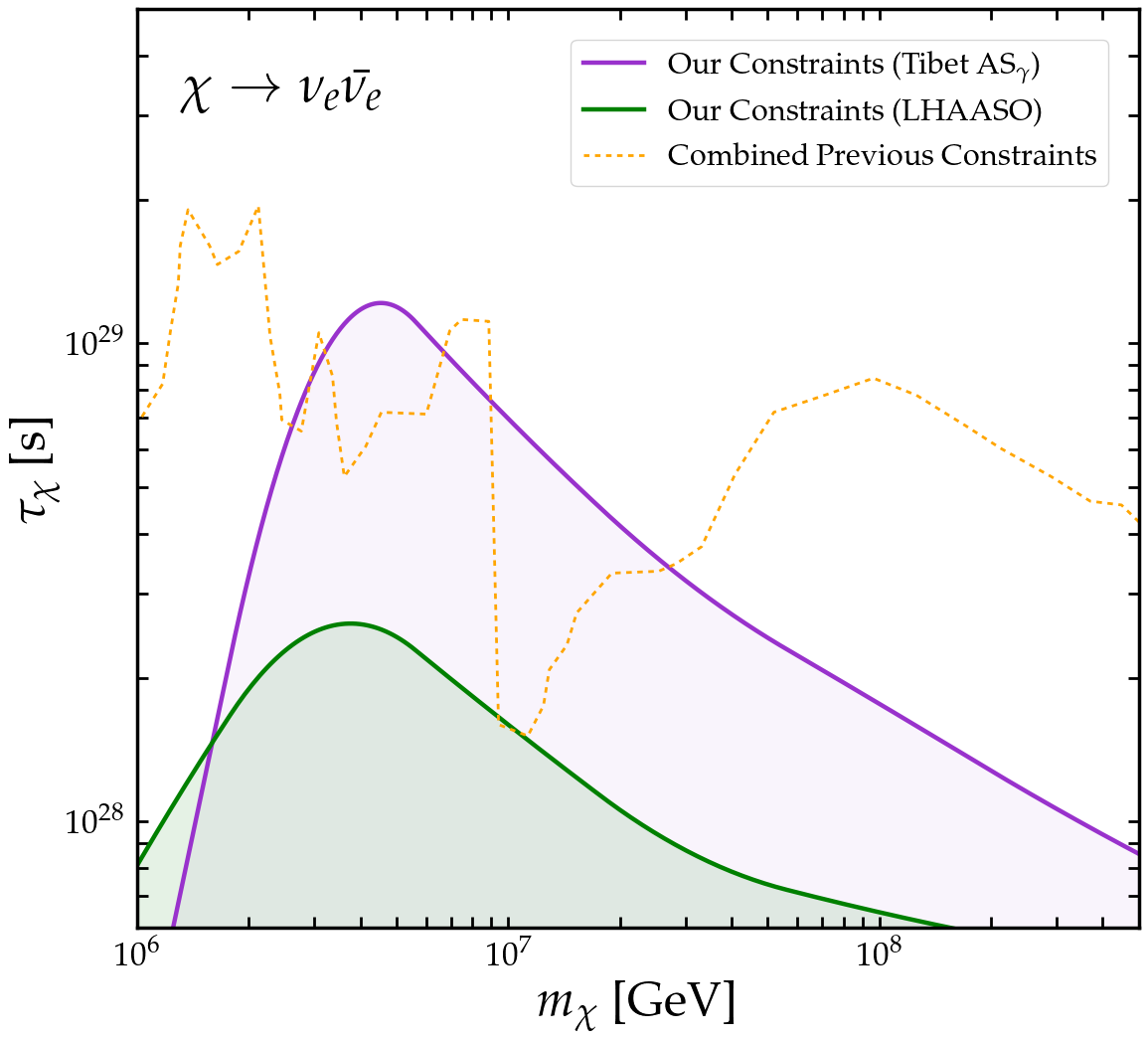} &
      \includegraphics[width=0.32\textwidth]{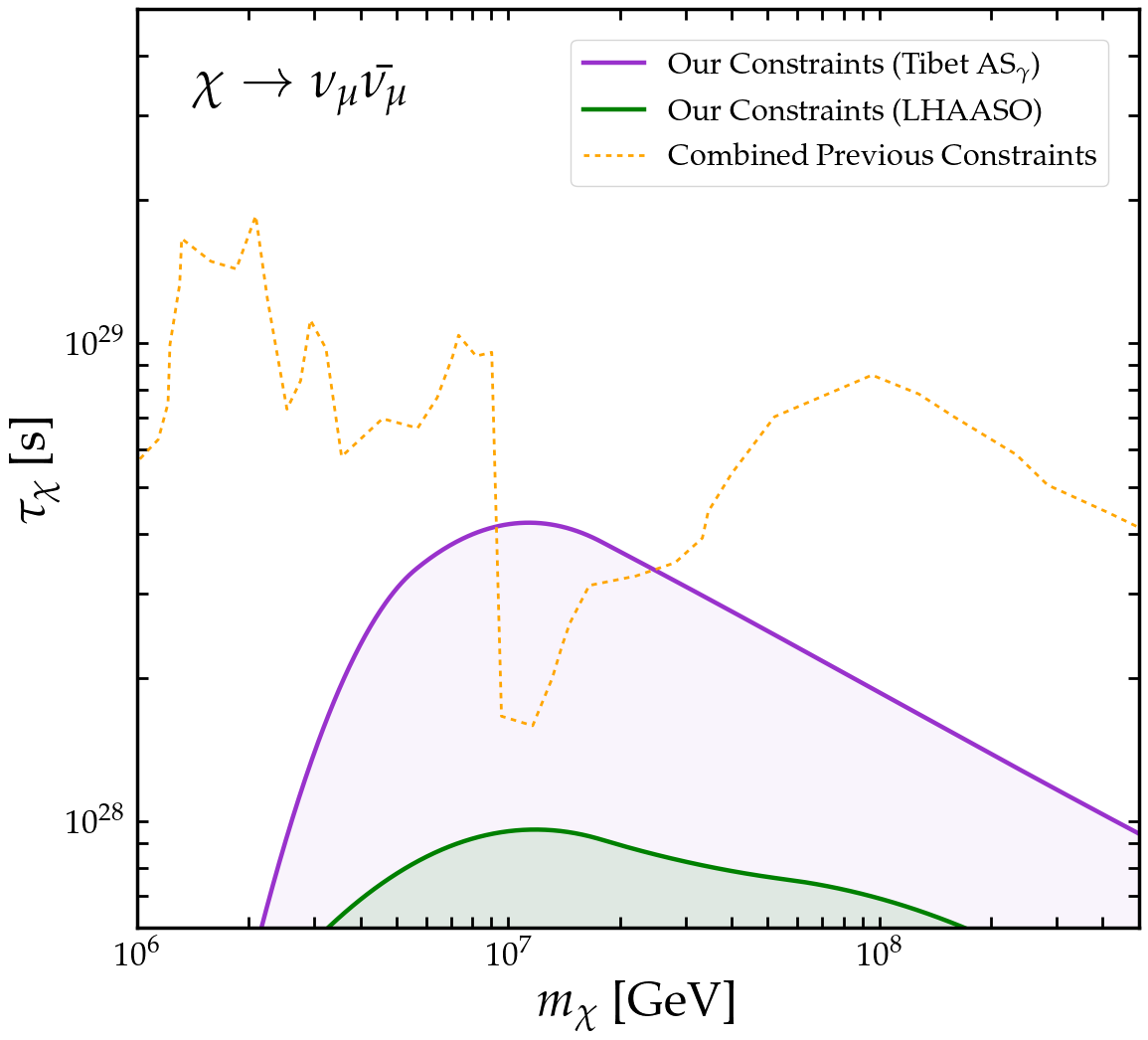} &
      \includegraphics[width=0.32\textwidth]{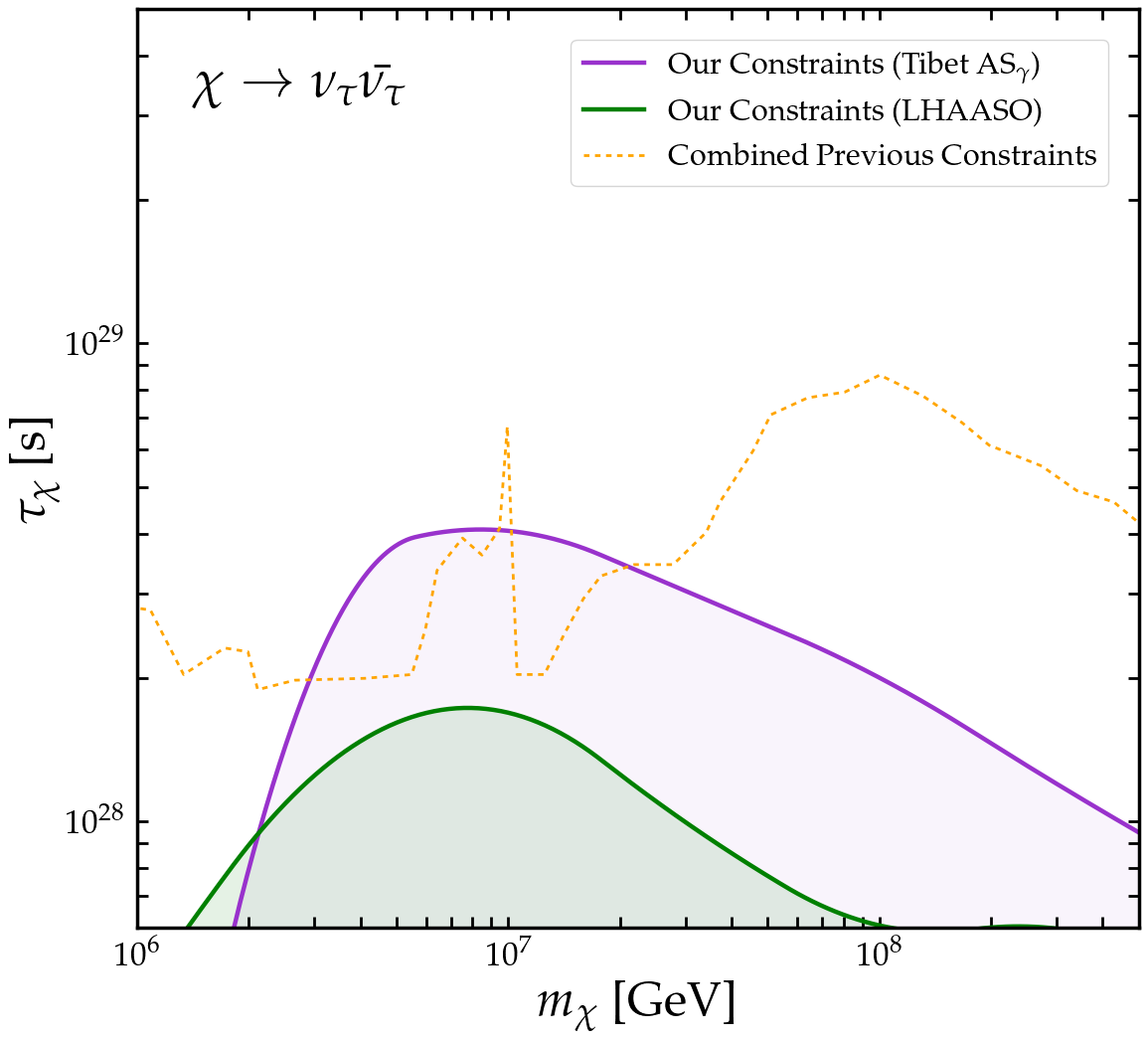} \\
    \end{tabular}%
  }

  \caption{Bounds on DM lifetime for different DM decay channels. Our bounds from Tibet AS$\gamma$ (Neronov \textit{et al.})\,\cite{Neronov:2021ezg} and LHAASO-KM2A\,\cite{LHAASO:2023gne} datasets are shown by the purple solid and green solid lines, respectively. Previous combined best bound in the parameter space is taken from Refs.\,\cite{Esmaili:2014rma,Cohen:2016uyg,Blanco:2018esa,IceCube:2018tkk,Bhattacharya:2019ucd,Ishiwata:2019aet,Chianese:2019kyl,LHAASO:2022yxw,LHAASO:2024upb} (orange dashed line). For ease of comparison, the $y$ axis range is kept the same across all the plots. }
\end{figure}


\end{document}